\documentclass[10pt,aps,prl,twocolumn,superscriptaddress]{revtex4-2}
\usepackage{dcolumn}
\usepackage{mathtools}
\usepackage{amssymb}
\usepackage{bm}
\usepackage{pifont}
\usepackage{psfrag}
\usepackage{epstopdf}
\usepackage{amsmath}
\usepackage{hyperref}
\usepackage{lineno}
\usepackage{lipsum}
\usepackage{comment}
\usepackage{siunitx}
\usepackage{graphicx}
\usepackage{setspace}
\hypersetup{
     colorlinks = true,
     linkcolor = blue,
     anchorcolor = blue,
     citecolor = blue,
     filecolor = blue,
     urlcolor = blue
     }
\usepackage[usenames]{color}
\renewcommand{\footnoterule}{%
  \hrule width \textwidth height 1pt
  \kern 2pt
}

\usepackage{float}

\newcommand{\suppsectionheading}[1]{%
  \par\addvspace{2.8ex}%
  \noindent{\fontsize{15}{18}\selectfont\bfseries #1\par}%
  \nobreak\vspace{1.2ex}%
}
\newcommand{\suppsubsection}[1]{%
  \par\addvspace{2.4ex}%
  \noindent{\fontsize{14}{17}\selectfont\bfseries #1\par}%
  \nobreak\vspace{0.8ex}%
}
\newcommand{\suppsubsubsection}[1]{%
  \par\addvspace{2.0ex}%
  \noindent{\fontsize{12}{14.4}\selectfont\bfseries #1\par}%
  \nobreak\vspace{0.6ex}%
}
\newcommand{\suppcaption}[1]{%
  \par\vspace{0.45\baselineskip}%
  \begingroup
  \fontsize{10.5}{12.5}\selectfont
  \setlength{\parindent}{0pt}%
  \raggedright
  \noindent #1\par
  \endgroup
  \vspace{0.45\baselineskip}%
}
\newcommand{\suppcite}[1]{\textup{[#1]}}

\makeatletter
\newcommand{\listsuppcontents}{%
    \suppsectionheading{Supplementary Contents}%
    \@starttoc{sup}%
}
\newcommand{\suppnote}[2]{%
    \phantomsection%
    \suppsectionheading{Supplementary Note #1 \textbar\ #2}%
    \addcontentsline{sup}{suppentry}{%
        \textbf{Supplementary Note #1} \textbar\ #2%
    }%
}
\newcommand{\suppfigentry}[2]{%
    \phantomsection%
    \addcontentsline{sup}{suppentry}{%
        \textbf{Supplementary Fig. #1} \textbar\ #2%
    }%
}
\newcommand{\suppvideosection}[1]{%
    \phantomsection%
    \suppsectionheading{Supplementary Videos \textbar\ #1}%
    \addcontentsline{sup}{suppentry}{%
        \textbf{Supplementary Videos} \textbar\ #1%
    }%
}
\newcommand*\l@suppentry{\@dottedtocline{1}{0em}{0em}}
\makeatother

\begin{document}
\nolinenumbers
\title{Geometry-Controlled Relaxation Spectra in Viscoelastic Fluids}
%\title{Geometry-induced proliferation of memory effects in  viscoelastic fluids}
\author{Niloyendu Roy}
\affiliation{Fachbereich Physik, Universität Konstanz, 78457 Konstanz, Germany}
\author{Rupayan Saha}
\affiliation{Institut für Theoretische Physik, Georg-August-Universität Göttingen,
37077 Göttingen, Germany}
\author{Debankur Das}
\affiliation{Institut für Theoretische Physik, Georg-August-Universität Göttingen,
37077 Göttingen, Germany}
\author{Matthias Kr{\"u}ger}
\affiliation{Institut für Theoretische Physik, Georg-August-Universität Göttingen,
37077 Göttingen, Germany}
\author{Clemens Bechinger}
\affiliation{Fachbereich Physik, Universität Konstanz, 78457 Konstanz, Germany}

\begin{abstract}
Soft materials store, dissipate and release mechanical stresses through relaxation processes that often span many orders of magnitude in time. Such relaxation spectra are widely used to infer internal material dynamics and are usually regarded as fingerprints of microscopic complexity, disorder, or heterogeneity. Here we show that a broad relaxation spectrum can instead be generated by the geometry of mechanical excitation itself. Using rotationally driven colloidal dimers in a wormlike micellar fluid with a dominant bulk relaxation time of order one second, we demonstrate that torsional driving converts distance from the driven object into relaxation time. This produces a geometry-controlled hierarchy of relaxation modes: orientational recoils persist for hundreds of seconds and encode past torque protocols over comparably long times. Particle velocimetry reveals rapid angular-momentum transport away from the probe, in contrast to the slow relaxation of stored torsional stress. A continuum shell model captures the observed recoil dynamics and the selective suppression of long-lived contributions under spatial confinement. Our results show that geometry can transform a material with simple intrinsic relaxation into a system with long-lived, space-dependent memory, suggesting a route to tune material dynamics through mechanical excitation rather than composition, with potential implications for microscopic mechanical memory elements.
\end{abstract}

%\date{\today}
\maketitle

\setlength\columnsep{25pt}

\section*{Introduction}

The response of materials to external perturbations is never instantaneous. Deformations and stresses generated by forcing typically persist after the perturbation has changed or ceased, causing the present state of a system to depend on its driving history. This history dependence constitutes memory, in the sense that earlier mechanical states leave a persistent imprint on the material’s present response~\cite{ferry1980viscoelastic}. Such effects are particularly pronounced in soft and complex materials, where relatively weak forces can induce substantial deformation or structural rearrangement, while the resulting stresses and structures may persist for long times. The recovery of these stresses and structures is commonly described by characteristic relaxation times or, more generally, by a relaxation spectrum, usually interpreted as a material-intrinsic fingerprint of microscopic and mesoscopic dynamics~\cite{larson2005rheology,chaudhuri2007reversible,zhou2018dynamically,kovacs1963glass,Mandal2021,Keim2019,zia2013stress,Vaidya2025}. Controlling this relaxation is central to soft-material technologies ranging from dissipative and adhesive materials to biomedical gels and soft robotic actuators~\cite{creton2016fracture,zhou2016viscoelastic,li2022soft}.

%\cite{song2016rheological,goudoulas2016viscoelastic,cates1990statics,larson2005rheology,chaudhuri2007reversible,ross1992structure,zhou2018dynamically,trepat2007universal,lieleg2009cytoskeletal,mofrad2009rheology,Mandal2021,kovacs1963glass,Keim2019,Lahini2017,Murphy2020,Vaidya2025}. 

Relaxation spectra, however, need not be inherited from intrinsic material modes alone. Even in Newtonian fluids, spatial transport can generate memory: a local perturbation spreads momentum through the surrounding medium, and the resulting flow can act back on the probe at later times, producing algebraically decaying translational and rotational correlations~\cite{paul1981observation,cichocki2000long,franosch2011resonances}. Viscoelastic media add a qualitatively different ingredient. Deformation generated near the probe can be transported through the material and stored as mechanical stress, so that rapid momentum transport becomes separated from slow stress relaxation. If the forcing geometry distributes and preserves this stored stress over spatially distinct regions, the response measured at the probe can acquire relaxation times generated by geometry rather than by bulk material properties.

Here we realize this mechanism experimentally with rotationally driven micron-sized colloidal dimers in a viscoelastic fluid whose bulk response is dominated by a single relaxation time of about one second. During such driving, the dimer sets up a torsional deformation within the medium, with material regions sheared around a fixed centre while retaining their radial ordering. From a Lagrangian perspective, material elements at a given radius remain part of the same deforming region and can therefore store torsional stress over extended times. This spatial organization converts radial distance into a specific relaxation time: stresses accumulated at different distances from the probe provide distinct relaxation contributions, which together form a hierarchy of timescales extending far beyond the single bulk relaxation time. Depending on the driving history, even torsional stresses of opposite sign can be stored at different radii; their sequential release then causes the recoil itself to reverse direction. A continuum shell model quantitatively captures this radius-dependent relaxation and predicts that confinement selectively suppresses the longest-lived modes, as confirmed experimentally. Because the effect relies on transport, preserved deformation geometry and local stress storage rather than on specific microscopic details, our results establish a general route to control long-lived material memory through geometry rather than composition. This suggests opportunities for microscopic information storage and processing based on mechanically encoded histories.
%Because the effect relies on transport, preserved deformation geometry and local stress storage rather than on specific microscopic details, our results establish a general route to control long-lived material memory through geometry rather than composition, with potential implications for microscopic soft-matter devices and mechanically encoded memory elements. This may open new perspectives for microscopic information storage and processing in soft and biological materials, for example in mechanically addressable memory elements, history-dependent microrheological sensors, rotary micromachines and viscoelastic cellular or extracellular environments where localized forcing encodes mechanical histories in spatially organized stress fields.

\begin{figure*}
\centering
\includegraphics[width=0.6507\linewidth]{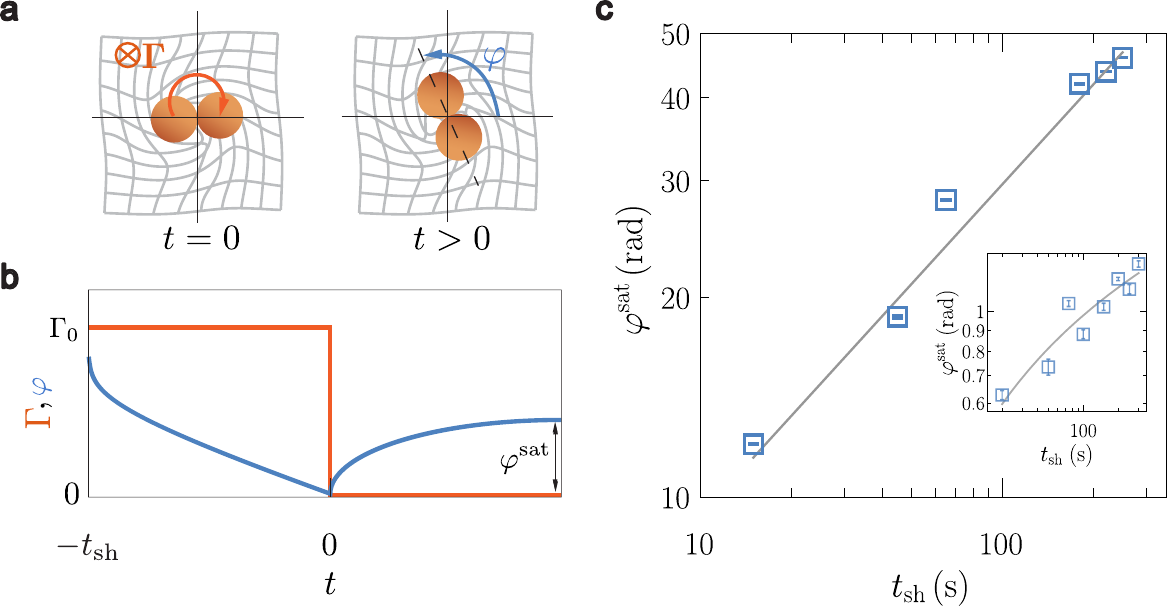}
\caption{\textbf{Torsional driving generates long-lived orientational recoil.}
\textbf{a} Schematic illustration of a viscoelastic network (gray) deformed by a driven colloidal dimer (orange). During $-t_{\rm sh}<t<0$, the applied torque $\Gamma$ rotates the dimer and distorts the surrounding micellar network, leading to stored torsional stress. After the torque is switched off at $t=0$, relaxation of this deformation drives an orientational recoil (blue arrow) opposite to the driving direction.
\textbf{b} Applied torque protocol and corresponding orientational response of the dimer. The recoil amplitude $\varphi^{\rm sat}$ is defined as the asymptotic angular displacement after torque removal.
\textbf{c} Recoil amplitude $\varphi^{\rm sat}$ measured as a function of driving time $t_{\rm sh}$ for an applied torque magnitude $\Gamma_0=192\,\mathrm{pN}\,\mu\mathrm{m}$. Symbols denote experimental data; the solid line indicates a power-law dependence, $\varphi^{\rm sat}\propto \sqrt{t_{\rm sh}}$. Inset, corresponding data for $\Gamma_0=7.2\,\mathrm{pN}\,\mu\mathrm{m}$, with the solid line showing a fit $\varphi^{\rm sat}\propto \ln t_{\rm sh}$. Error bars indicate the standard error of the mean from 15 independent experiments.}
\label{Unbounded_Large_Recoil}
\end{figure*}

\section*{Experimental Setup}

In our experiments, we use an equimolar aqueous solution of cetylpyridinium chloride monohydrate (CPyCl) and sodium salicylate (NaSal) at a concentration of $5.6$ mM and temperature $298$ K. Under these conditions, the fluid forms an entangled network of wormlike micelles whose bulk stress relaxation is approximately single exponential, with characteristic time $\tau_{\mathrm{b}}\approx1$ s \cite{cates1990statics,baiesi2021rise,sung2003rheological,SpenleyCates1993}. Microscopic recoil experiments reveal an additional faster timescale associated with colloid--fluid coupling \cite{gomez2014probing,caspers2023mobility}. Thus, under conventional translational driving, the fluid response is characterized by only a small number of relaxation modes.

As probes, we use colloidal dimers composed of two irreversibly linked superparamagnetic spheres with diameter $4.5~\mu$m (see Methods). Compared to a single spherical particle, a rotating dimer is substantially more effective at generating rotational flow fields; in addition it allows for precise detection of its orientation.  Due to their weight, the dimers sediment towards the bottom of the sample cell, where their two-dimensional motion is recorded by video microscopy.

Torsional driving is imposed by a spatially uniform rotating magnetic field $\mathbf{H}$ in the $x-y$ plane \cite{wilhelm2003rotational}
\begin{equation}
\mathbf{H}(t)
=
H
\begin{pmatrix}
\cos(\omega_{\mathrm{H}} t)\\
\sin(\omega_{\mathrm{H}} t)
\end{pmatrix}
\end{equation}
with fixed angular frequency $\omega_{\mathrm{H}}=20\pi~\mathrm{rad~s^{-1}}$ and time $t$. Owing to the phase lag between the induced magnetic moment $\mathbf{M}$ and the external field $\mathbf{H}$, the dimers experience a constant magnetic torque $\boldsymbol{\Gamma}=\mathbf{M}\times\mathbf{H}$ \cite{janssen2009controlled}.

\section*{Results}

We first probe how rotational driving stores and releases torsional stress in the micellar fluid. A colloidal dimer is driven at constant torque with magnitude $\Gamma_0$ for a prescribed time $t_{\mathrm{sh}}$ and then released, while its orientation $\varphi(t)$ is monitored after torque removal (Fig.~\ref{Unbounded_Large_Recoil}a,b; see Supplementary Fig.~S1 for experimental trajectories). During this interval, the dimer rotates about a fixed centre and sets up a torsional deformation of the medium.
Since the subsequent motion occurs without applied torque, this recoil protocol provides a direct readout of torsional stresses stored in the surrounding fluid during the driving phase. Upon torque removal, the dimer rotates opposite to the preceding driving direction. This orientational recoil shows that part of the work performed during driving is stored in the fluid and released after the external torque has ceased. Notably, for sufficiently long driving times, the recoil amounts to several complete turns of the dimer (Supplementary Video~1). We quantify this response by the asymptotic recoil amplitude $\varphi^{\mathrm{sat}}=\varphi(t\rightarrow\infty)$, defined as the total angular displacement occurring after the torque is switched off.

The recoil is driven by the relaxation of stress stored in the surrounding micellar network. A natural local interpretation would be that the rotating dimer winds nearby micellar strands, which then pull the dimer back after torque removal. We first test whether such a local storage picture can account for the observed recoil. In this picture, an angular displacement $\varphi^{\mathrm{sat}}$ requires a contour displacement of at least $l_{\min}\sim R\varphi^{\mathrm{sat}}$, where $R$ is the size of the rotating probe. The resulting values exceed $200\,\mu\mathrm{m}$, far larger than the typical contour lengths of individual wormlike micelles, which range from approximately $100\,\mathrm{nm}$ to $10\,\mu\mathrm{m}$~\cite{SpenleyCates1993,rehage1988rheological}. Since wormlike micelles are living polymers that continuously break and recombine, their contour lengths represent dynamical rather than permanent structural length scales. A purely local storage mechanism is therefore difficult to reconcile with the observations. Instead, the large orientational recoil indicates that torsional stress is stored over distances far exceeding the size of individual micellar constituents.

\begin{figure}
\centering
\includegraphics[width=0.85\columnwidth]{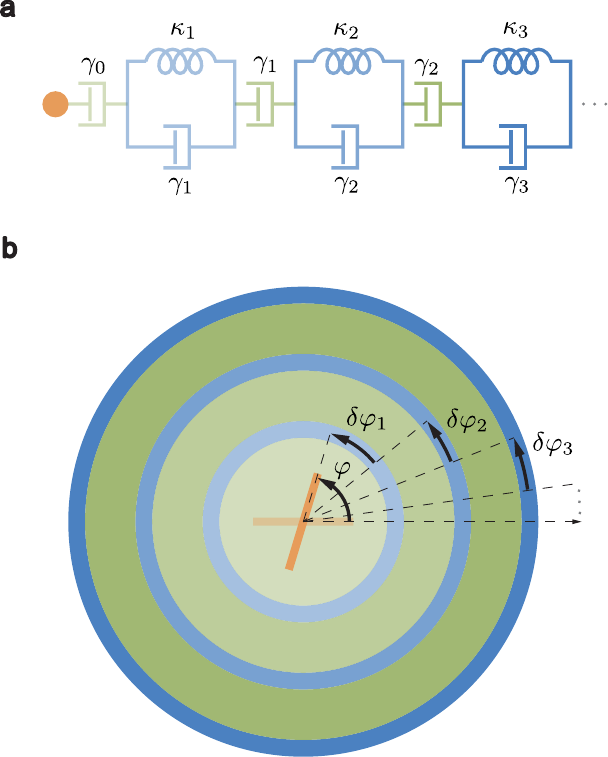}
\caption{\textbf{A shell model converts radial distance into relaxation time.}
\textbf{a} One-dimensional mechanical analogue of the shell model. The orange node denotes the driven dimer at the free left boundary. Neighbouring nodes are alternatingly coupled viscously by dashpots (green) or by  viscoelastic elements consisting of a spring and dashpot in parallel (blue). These viscous and viscoelastic couplings form a chain of Jeffrey-type units. The colour gradient indicates increasing damping $\gamma_i$ and stiffness $\kappa_i$ along the chain. Viscous couplings transmit momentum along the chain, whereas viscoelastic elements store and relax stress.
\textbf{b} Concentric-shell representation of the rotational geometry, with solid and faded orange directors at the centre indicating the dimer orientations at $t=0$ and $t>0$, respectively. Concentric green and blue shells represent viscous couplings and viscoelastic elements, respectively. Angular momentum is transported radially via the viscous couplings, whereas each viscoelastic element stores torsional stress and contributes an angular displacement $\delta\varphi_i$ during relaxation, as indicated by curved arc segments. The orientational recoil of the dimer is thus obtained as the sum $\varphi=\sum_i\delta\varphi_i$, consistent with Eq.~\eqref{rf_dimer}.}
\label{Fig_model}
\end{figure}

The dependence of the recoil amplitude on driving time provides a second indication of spatially distributed stress storage. Figure~\ref{Unbounded_Large_Recoil}c shows that $\varphi^{\mathrm{sat}}$ increases with driving time $t_{\mathrm{sh}}$ without any indication of saturation. Torsional stresses therefore continue to accumulate over times far beyond the bulk relaxation time, $\tau_{\mathrm{b}}\sim1\,\mathrm{s}$. Depending on the driving conditions, the growth in Fig.~\ref{Unbounded_Large_Recoil}c is well described by either a power law or a logarithm, indicating the absence of a single dominant relaxation time. This behaviour is qualitatively different from translational recoil in the same fluid, where changing the driving protocol affects only the response amplitudes while the characteristic relaxation times remain essentially unchanged (Supplementary Note~1 and Fig. S2)~\cite{ginot2022recoil,caspers2023mobility}. The lack of saturation in the orientational recoil therefore does not simply reflect stronger loading of pre-existing material relaxation modes; instead, it points to a rotationally generated hierarchy of spatially distributed stress-storage modes.

To describe how rotational geometry converts spatially distributed stress storage into a relaxation spectrum, we introduce a shell model in which the medium surrounding the probe is represented by mechanically coupled concentric shells. Each shell can store and relax torsional stress, while neighbouring shells transmit angular momentum. The model is based on two experimentally motivated ingredients: torsional stresses are distributed over the surrounding medium rather than confined to the immediate vicinity of the probe, and shells farther from the probe relax on progressively longer timescales. This provides a minimal framework in which radial position is converted into relaxation time, allowing torsional driving to generate a broad spectrum of relaxation modes.

Figure~\ref{Fig_model}a shows a one-dimensional mechanical analogue of the shell model, using standard spring--dashpot representations of linear viscoelasticity~\cite{ferry1980viscoelastic}. In this analogue, the free left end of the chain represents the driven dimer, while the fixed right boundary represents the outer system boundary. Neighbouring nodes are coupled by viscous dashpots, and every other pair of nodes is additionally connected by a spring element. The resulting chain can be viewed as a spatially extended sequence of Jeffrey-type units: viscous couplings permit relative motion between neighbouring regions and establish the steady velocity profile, whereas the spring--dashpot elements store and relax stress. When a force is applied at the free boundary, marked by the orange node, momentum is transmitted through the viscous couplings, while stress is stored and subsequently released by the viscoelastic elements. Compared with a single Jeffrey-type element, the chain therefore introduces spatially distributed stress storage and generates a spectrum of relaxation modes.

To translate this mechanical analogue into the rotational geometry of the experiment, we map the linear chain onto a set of \(N\) concentric shells surrounding a dimer rotating about a fixed centre (Fig.~\ref{Fig_model}b). The force applied at the free end of the chain  thereby turns into a torque acting on the dimer, schematically indicated by the orange bar in Fig.~\ref{Fig_model}b. Adjacent shells retain their radial ordering and are coupled viscously, so that angular momentum is transported radially away from the probe, analogous to momentum transport along the dashpot-coupled chain. The spring--dashpot elements of the one-dimensional model become viscoelastic degrees of freedom associated with the individual shells, where they store and relax torsional stress.  This mapping separates angular-momentum transport from torsional-stress relaxation and makes radial distance a variable controlling the stored stress and its relaxation time. The equations of motion for the model governing shell dynamics are given in Eq.~(1) and detailed in  Fig.~S3 of Supplementary Note~2.

The solution of the model \cite{Guo2004EigenViscoelastic,SerraAguila2019,saha2026peculiar} has three experimentally useful consequences. First, each viscoelastic unit in Fig.~\ref{Fig_model}a corresponds to an eigenmode of the chain, with relaxation time $\tau_i=\gamma_i/\kappa_i$. Second, after the torque is switched off, %following a long driving time $t_{\mathrm{sh}}\rightarrow\infty$, 
the recoil is governed by the stress stored in the viscoelastic units. For $t_{\mathrm{sh}}\rightarrow\infty$, shell $i$ contributes an angular displacement
\begin{align}
\delta\varphi_i(t)=\frac{\Gamma_0}{\kappa_i}\left(1-e^{-\kappa_i t/\gamma_i}\right).\label{eq:deltaphi}
\end{align}
The dimer recoil is given by $\varphi(t)=\sum_i\delta\varphi_i(t)$ and the asymptotic recoil amplitude is $\varphi^{\mathrm{sat}} = \Gamma_0\sum_i 1/\kappa_i $. Third, the steady-state angular-velocity profile under applied torque is independent of the viscoelastic units and is determined by the viscous couplings alone. The model therefore separates angular-momentum transport from torsional-stress storage, giving the parameters $\gamma_i$ and $\kappa_i$ direct and experimentally testable meanings.

These properties can be summarized by the linear-response relation
\[
\varphi(t)=\int_{-\infty}^{0}\mathrm{d}t'\,\chi(-t')\,\Gamma(t')-\int_{-\infty}^{t}\mathrm{d}t'\,\chi(t-t')\,\Gamma(t'),
\]
with linear response function (susceptibility), defined in Supplementary Note 2,
\begin{equation}
\chi(t)=
\underbrace{\left(\frac{1}{\gamma_0}+\sum_{i=1}^{N}\frac{1}{\gamma_i}\right)}_{\mathrm{viscous}}
+
\underbrace{\sum_{i=1}^{N}\frac{1}{\gamma_i} e^{-\frac{\kappa_i}{\gamma_i} t}}_{\mathrm{viscoelastic}}.
\label{rf_dimer}
\end{equation}
Equation~\eqref{rf_dimer} makes the separation between angular-momentum transport and torsional-stress relaxation explicit. The time-independent terms describe the instantaneous viscous response that sets the long-time angular velocity under a constant applied torque, whereas the exponential terms describe delayed viscoelastic relaxation after torque removal. Because each shell enters independently through its parameters $\gamma_i$ and $\kappa_i$, the recoil is represented as a superposition of spatially distributed relaxation modes. A detailed derivation is given in Supplementary Note~2, and the shell dynamics are visualized in Supplementary Video~2.

To identify how rotational geometry generates the relaxation spectrum, we take the continuum limit of the shell model. The shell radius is promoted to a continuous radial coordinate, $r\in(R,L)$, where $R$ denotes the probe size and $L$ the outer system size. The discrete parameters $\gamma_i$ and $\kappa_i$ then become continuous functions $\gamma(r)$ and $\kappa(r)$. This limit is obtained by taking $N\rightarrow\infty$ and $\Delta r\rightarrow 0$, while keeping $L-R=N\Delta r$ fixed \footnote{The boundary conditions turn into a free boundary condition at $r=R$ and a Dirichlet boundary condition at $r=L$. $\Delta r$ is the radial distance between neighboring shells.}. The response function, Eq.~\eqref{rf_dimer}, becomes
\begin{equation}
\chi(t)=\int_R^L \frac{\mathrm{d}r}{\gamma(r)}
\left[1+\exp\left(-\frac{t}{\tau(r)}\right)\right].
\label{eq_continuum_response}
\end{equation}
This continuum form makes the geometric origin of the relaxation hierarchy explicit: each radial position contributes a relaxation time $\tau(r)=\gamma(r)/\kappa(r)$.  The angular velocity $\dot\varphi_i$ of shell $i$ turns into the continuous angular velocity denoted $\omega(r)$, by definition corresponding to steady state (Supplementary Note~2).

A key consequence of the continuum description is that the hierarchy of relaxation times emerges directly from the radial geometry of the flow. To see this, we determine how transport and stress storage scale with radial distance. Geometric surface-area scaling (see Eq.~(14) in Supplementary Note~2) implies that $\gamma(r)\sim r^{4}$ for spherical shells, or,  $\gamma(r)\sim r^{3}$ for cylindrical shells. Equivalently, for dimensionality $d=3$ for spheres and $d=2$ for cylinders, this yields $\gamma(r)\sim r^{d+1}$. 
%For axisymmetric Stokes flow in $d$ spatial dimensions, the viscous coupling scales as $\gamma(r)\sim r^{d+1}$. 
The steady-state angular velocity of the fluid at radial distance $r$ therefore scales as (taking $L\gg r$)
\begin{equation}
\omega(r)\sim r^{-d},
\end{equation}
in agreement with Stokes flow for  a rotating sphere or cylinder~\cite{landau1987fluid}. From Eq.~\eqref{eq:deltaphi}, the elastic coefficient $\kappa(r)$ sets the relaxation amplitude associated with a shell at distance $r$, which scales as $1/\kappa(r)$. We assume that this relaxation corresponds to an $r$-independent microscopic contour displacement (e.g., associated with a micellar length scale), so that the angular displacement scales as $\sim 1/r$ \footnote{This is because arc length is radius multiplied by covered angle.}. This gives $1/\kappa(r)\sim 1/r$ and hence $\kappa(r)\sim r$. The relaxation time associated with radius $r$ is therefore
\begin{equation}
\tau(r)=\frac{\gamma(r)}{\kappa(r)}
\sim r^d .
\end{equation}
The shell model thus predicts a striking separation between amplitude and timescale. Regions farther from the dimer contribute progressively smaller recoil amplitudes, scaling as $1/r$, but relax over progressively longer times, scaling as $r^d$. A broad hierarchy of relaxation times therefore emerges even when the local viscoelastic response is characterized by a single microscopic length scale. Long-lived memory is not imposed by an intrinsic spectrum of material relaxation modes; instead, it is generated by the rotational geometry of the deformation, which converts radial position into relaxation time and is cut off only by the system size.

We now focus on $d=3$, corresponding to spherical shells around the rotating probe, as appropriate to our experiments.
In the unconfined limit, $L\rightarrow\infty$, Eq.~\eqref{eq_continuum_response} gives
\begin{align}
\chi(t)-\chi(t\rightarrow\infty) \sim \frac{1}{t}.\label{eq:1t}
\end{align}
Because the recoil is obtained by integrating this delayed part of the response over time, the angular displacement grows as $\varphi(t)\sim \ln t$ for $t_{\mathrm{sh}}\rightarrow\infty$ (Supplementary Note~2). For finite driving time, Eq.~\eqref{eq:1t} implies a saturation amplitude $\varphi^{\mathrm{sat}}\sim\ln t_{\mathrm{sh}}$. The logarithmic dependence observed in Fig.~\ref{Unbounded_Large_Recoil}c, inset, therefore directly supports the predicted geometry-generated hierarchy of relaxation times. 

\begin{figure*}
\centering
\includegraphics[width=0.90\linewidth]{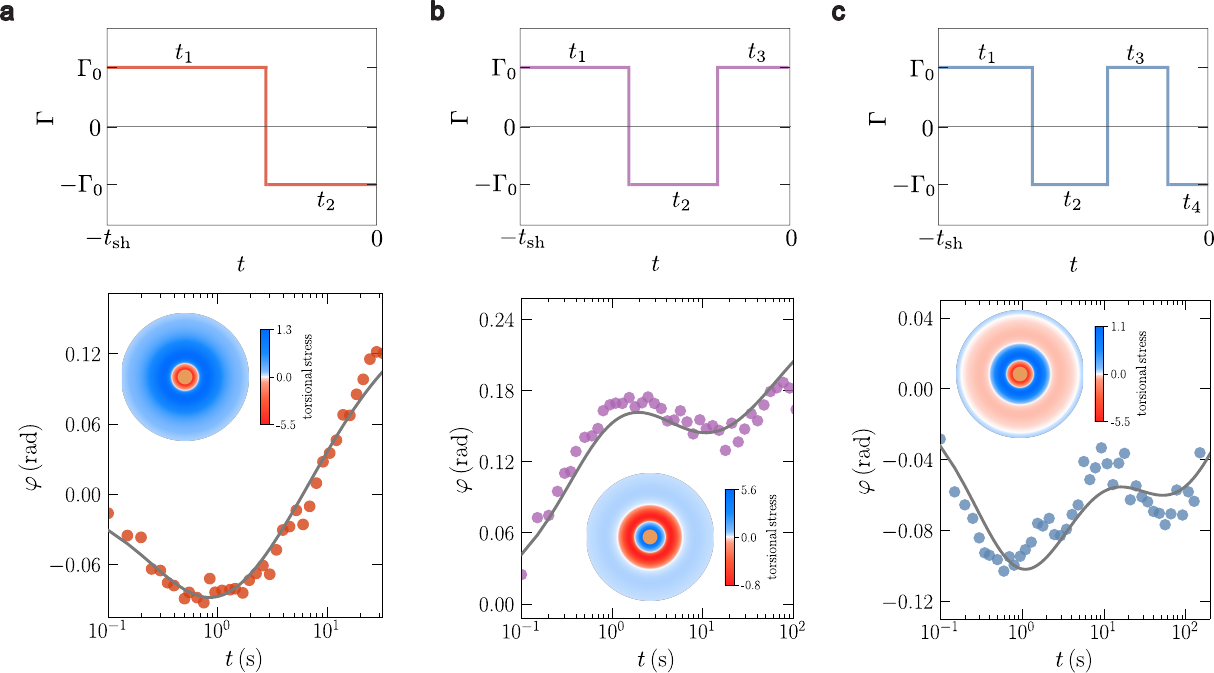}
\caption{\textbf{Torque-reversal protocols encode memory in orientational recoil.}
\textbf{a--c} Recoil response after torque protocols with one, two and three reversals, respectively. Top panels show the applied time-dependent torque protocols (note that time periods are not to scale), in which the torque is switched between $\pm\Gamma_0$ for successive durations indicated in each panel. In \textbf{a}, \textbf{b} and \textbf{c}, $(t_1,t_2)=(20,0.8)\,\mathrm{s}$, $(t_1,t_2,t_3)=(85,7.5,1.2)\,\mathrm{s}$, and $(t_1,t_2,t_3,t_4)=(600,50,5,0.6)\,\mathrm{s}$, respectively. Bottom panels show the corresponding orientational recoil of the dimer after the torque is switched off. In each case, the recoil trajectory retains the history of the preceding driving sequence: one, two and three torque reversals lead to one, two and three reversals in the subsequent recoil, respectively. Symbols denote experimental averages from 15 independent experiments; grey curves are theoretical fits from the shell model, with slightly different driving intervals in \textbf{b} and \textbf{c}  (see Supplementary Note 4 and 5 for details and for fitting parameters).
Insets show radial profiles of the torsional stress at $t=0$ obtained from the shell model. The yellow circle mimics the dimer, and colour-coded regions represent the stress in the fluid between $r=7\,\mu\mathrm{m}$ and $r=60\,\mu\mathrm{m}$, indicating its alternating sign as a function of $r$.}
\label{fig_nonmonotinic}
\end{figure*}

The emergent hierarchy of relaxation modes gives the fluid a memory of the sequence in which torsional stress was applied. To test this memory experimentally, we use multi-step torque protocols in which the sign of the applied torque is repeatedly reversed between $+\Gamma_0$ and $-\Gamma_0$, with ${\Gamma}_0=7.2\,\mathrm{pN}\,\mu\mathrm{m}$ (Fig.~3a--c, top panels). The sign of $\Gamma$ denotes the sense of rotation. Because the magnetic-field reversals occur on timescales much shorter than the dimer dynamics, the corresponding torque changes can be considered effectively instantaneous (Supplementary Note~3 and Fig.~S4).

The resulting recoil responses are shown in the lower panels of Fig.~\ref{fig_nonmonotinic}. For protocols with one, two or three torque reversals, the dimer changes its recoil direction once, twice or three times, respectively, in  agreement with the shell model (lines). This one-to-one correspondence shows that the material does not merely retain the final state of the driving protocol, but encodes the preceding torque sequence. In the shell model, each torque interval loads different parts of the radial stress profile, so that torsional stresses of alternating sign can be stored at different radial distances at $t=0$ (insets in Fig.~\ref{fig_nonmonotinic}; see Supplementary Note~4). After release, these stored stresses relax sequentially and generate the observed recoil reversals (See Supplementary Videos~3--4 for visualization through the shell picture). Such non-monotonic recoil cannot be reduced to a single slow relaxation process; it requires a hierarchy of modes that store and release different parts of the applied torque history. Additional examples at higher torque, $\Gamma_0=192\,\mathrm{pN}\,\mu\mathrm{m}$, are provided in Supplementary Videos~5--7, Supplementary Note~6 and Fig.~S5.

This memory effect is robust across experimental conditions and material systems. Similar non-monotonic recoils are observed both near the bottom surface and in the bulk of the sample (Supplementary Note~7 and Fig.~S6). Qualitatively similar behaviour is also found in polyacrylamide (PAAM) solutions (Supplementary Note~8 and Fig.~S7). These observations indicate that the geometry-controlled hierarchy of relaxation times does not rely on microscopic details of the wormlike micellar system, but reflects the generic coupling of torsional geometry to delayed stress relaxation in viscoelastic media.

\begin{figure}
\centering
\includegraphics[width=\linewidth]{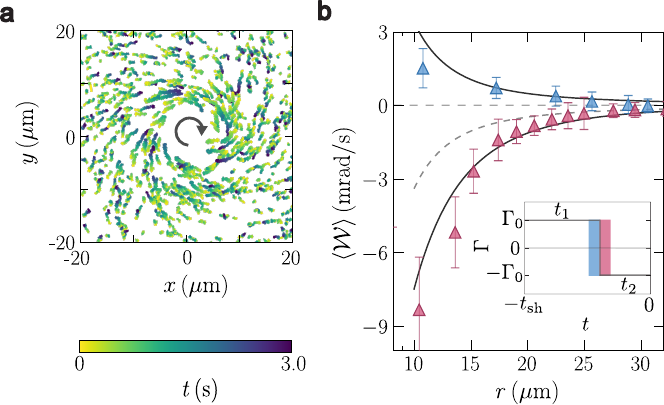}

\caption{\textbf{Tracer-particle velocimetry shows rapid angular-momentum transport.}
\textbf{a} Trajectories of non-magnetic silica particles of diameter $0.7\,\mu\mathrm{m}$, recorded during steady-state driving with $\Gamma_0=192\,\mathrm{pN}\,\mu\mathrm{m}$. The dimer centre of mass is located at the origin, and the driving direction is indicated by the arrow. Each trajectory is colour-coded by time.
\textbf{b} Blue and red triangles show the mean azimuthal velocity $\langle\mathcal{W}\rangle$ of the tracer particles as a function of radial distance $r=\sqrt{x^2+y^2}$, calculated over intervals $\Delta t=0.1\,\mathrm{s}$ indicated by the blue and red shaded regions in the inset. The inset schematically shows the driving protocol, defined by $(t_1,t_2)=(5,1)\,\mathrm{s}$ and $\Gamma_0=192\,\mathrm{pN}\,\mu\mathrm{m}$. Here, $\mathcal{W}=\Delta\phi/\Delta t$, with tracer azimuthal coordinate $\phi=\arctan(y/x)$, and $\langle\cdot\rangle$ denotes the average over all tracers at distance $r$. Error bars indicate the standard error of the mean from 20 independent experiments. Solid curves are theoretical predictions from the shell model. The dashed curve is the reflection of the theoretical prediction during driving about the line $\langle\mathcal{W}\rangle=0$.}

\label{Tracer_Flip}
\end{figure}

A further consequence of the shell model is the separation between fast angular-momentum transport and torsional stress relaxation. The broad hierarchy of relaxation times arises from the latter, whereas angular momentum is transmitted rapidly through viscous couplings, as represented by the instantaneous viscous response in Eq.~\eqref{rf_dimer}. Consequently, memory should be visible in slow observables such as orientational recoil, yet remain largely hidden in fast observables such as flow velocities. To test this separation, we use non-magnetic tracer particles to image the flow field around the dimer; representative trajectories are shown in Fig.~\ref{Tracer_Flip}a. Figure~\ref{Tracer_Flip}b shows the tracer angular velocities as a function of radial position $r$, measured immediately before and after a torque reversal, as sketched in the inset of Fig.~\ref{Tracer_Flip}b. The sign of the tracers' angular velocity reverses almost simultaneously over the entire measured range of $r$ (Supplementary Video~8), in agreement with the model calculations (solid lines). This demonstrates that angular momentum is transported on timescales far shorter than those governing orientational recoil, 
confirming the predicted separation between rapid transport and slow stress relaxation.

In addition to the sign reversal of the tracers' angular velocity, the torque reversal transiently enhances its magnitude, which is quantitatively captured by the shell model. This effect reflects the release of stored elastic stress, which adds to the viscous angular-momentum transport immediately after torque reversal. Tracer-particle velocimetry therefore not only captures the fast transport of angular momentum through dissipation, but also bears signatures of the stored elastic energy. The effect of torsional stress relaxation can also be seen by imaging tracer particles \emph{after} torque has been switched off (Supplementary Note~9, Fig.~S8 and Supplementary Video~9), where the tracers' angular motion itself exhibits non-monotonic behaviour during non-monotonic recoil of the dimer.

\begin{figure*}
\centering
\includegraphics[width=0.65\linewidth]{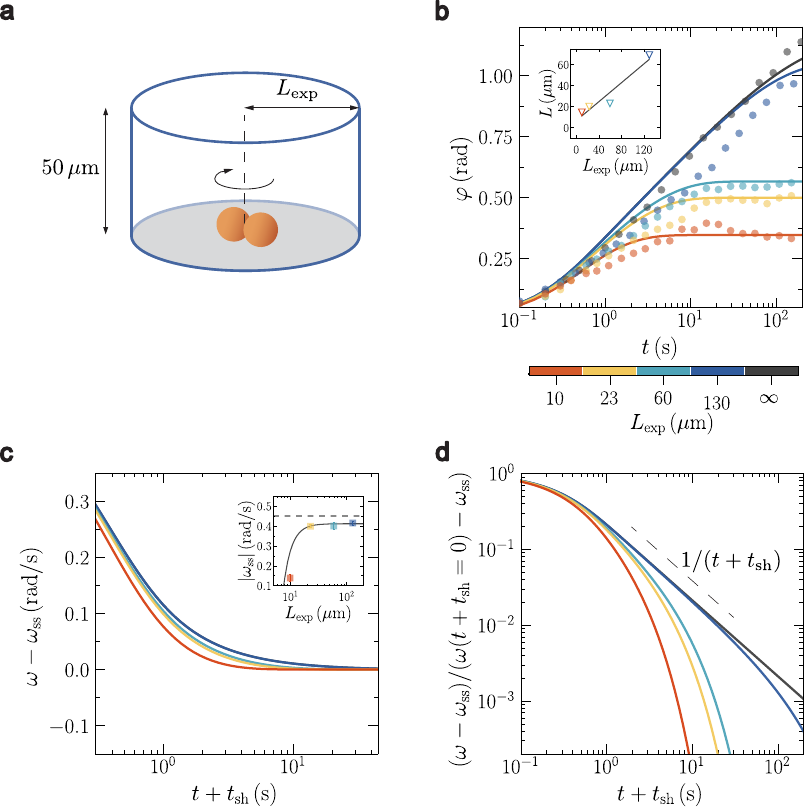}
\caption{\textbf{Confinement truncates the geometry-controlled relaxation spectrum.}
\textbf{a} Three-dimensional schematic of a cylindrical well of radius $L_{\mathrm{exp}}$ and depth $50\,\mu\mathrm{m}$, fabricated from SU-8 photoresist. The dimer sits on the bottom glass surface.
\textbf{b} Circles show the mean orientational recoil trajectory $\varphi(t)$ of the dimer inside confining wells of different radii $L_{\mathrm{exp}}$, as indicated by the colour bar, for $t_{\mathrm{sh}}=200\,\mathrm{s}$ and $\Gamma_0=7.2\,\mathrm{pN}\,\mu\mathrm{m}$. Each data point is averaged over 15 independent experiments. Solid lines are shell-model predictions using a model length scale $L$. Inset: Triangles show the fitted values of $L$ as a function of $L_{\mathrm{exp}}$; the grey line is a linear fit.
\textbf{c} Shell-model prediction for the angular frequency $\omega$ of the dimer, with the steady-state value $\omega_{\mathrm{ss}}$ subtracted, as a function of elapsed time after torque switch-on, $t+t_{\mathrm{sh}}$, for various confinement radii $L$, colour-coded as in \textbf{b}. Inset: Experimentally determined steady-state angular frequencies as a function of $L_{\mathrm{exp}}$ (data points) together with the shell-model prediction (solid line). The horizontal dashed line indicates the experimental value for the unconfined dimer. Error bars indicate the standard error of the mean from five independent experiments.
\textbf{d} Normalized shell-model curves corresponding to \textbf{c}, shown on logarithmic axes. The dashed line indicates the asymptotic $1/(t+t_{\mathrm{sh}})$ behaviour for $L\to\infty$.}
\label{fig_Confinement}
\end{figure*}

The shell model also suggests a direct test of the geometric origin of the relaxation hierarchy.
Since the largest relaxation times arise from the most distant regions of the fluid, limiting the radial extent available for stress storage should progressively remove the longest-lived components of the response. Confinement therefore provides a direct experimental test of whether the observed memory is set by geometry. We test this idea by confining the fluid within cylindrical wells of radius $L_{\mathrm{exp}}$, fabricated by photolithography (Methods), each with a height of approximately \(50\,\mu\mathrm{m}\) (Fig. ~\ref{fig_Confinement}a). Torque is applied while the dimer resides near the centre of the bottom surface of the well. Because the dimer position is not actively controlled, we restrict the analysis to trajectories for which its centre of mass remains close to the well centre, thereby isolating the radially symmetric effect of confinement.

Figure~\ref{fig_Confinement}b shows the orientational recoil following torsional driving for a fixed driving time \(t_{\mathrm{sh}}=200\,\mathrm{s}\) and different confinement radii {$L_{\mathrm{exp}}$} %\(r_B = L \) from the model, 
together with the unconfined case. Reducing the confinement radius progressively suppresses the recoil, demonstrating that limiting the radial extent of accessible fluid limits the amount of torsional stress that can be stored. The model predictions are shown as solid lines, with  good agreement. Minor deviations may partly be attributed to the fact that the experimental confinement is cylindrical and not spherical, as assumed in the model. The fitted model length scale $L$ is proportional to the experimental length scale $L_{\mathrm{exp}}$ (inset Fig.~\ref{fig_Confinement}b). Confinement truncates the relaxation hierarchy by removing the most distant, longest-lived modes, yielding a largest relaxation time of $\tau_{\rm max}=\tau_{\rm \min}\left ( \frac{L}{R}\right )^3$, with  $\tau_{\rm \min}$ the smallest relaxation time of the model. The curves thus saturate for $t\gtrsim \tau_{\rm max}$. The agreement of experimental and theoretical curves in Fig.~\ref{fig_Confinement}b directly supports the prediction that torsional memory is encoded non-locally through a geometric hierarchy of relaxation modes extending into the surrounding fluid. 

Even the unconfined recoil curve in Fig.~\ref{fig_Confinement}b saturates at long times. This saturation, however, reflects the finite driving duration rather than a cutoff imposed by system size. For $L\to\infty$, recoil follows $\sim\ln\frac{tt_{\rm sh}}{(t+t_{\rm sh})\tau_{\rm min}}$ (see Supplementary Note 2). This behaves, as already noted, as $\sim \ln t $ for $t\ll t_{\rm sh}$ and  $\sim \ln t_{\rm sh}$ for $t\gg t_{\rm sh}$, i.e., displaying saturation at $t \gtrsim t_{\rm sh}$. 

Figure~\ref{fig_Confinement}c shows the model prediction for the dimer's angular velocity with the steady-state value subtracted, as a function of time after torque startup, \(t+t_{\rm sh}\), for different values of $L_{\mathrm{exp}}$. In contrast to the recoil amplitude, the velocity depends only weakly on confinement and is therefore difficult to resolve experimentally. This weak dependence has a simple origin: recoil is an accumulated angular displacement, whereas velocity probes only the instantaneous rate of the viscoelastic response.
Thus, a logarithmically growing recoil corresponds to an approximately \(1/(t+t_{\rm sh})\) velocity response. Confinement cuts off this slow tail at the largest relaxation time, but this cutoff appears only at long times and is therefore difficult to extract from experimental velocity data. We therefore compare the experimentally accessible steady-state angular velocity \(\omega_{\mathrm{ss}}\), shown in the inset of Fig.~\ref{fig_Confinement}c. This steady-state velocity varies only weakly with $L_{\mathrm{exp}}$, in agreement with the shell model (solid line), which approaches the unconfined limit as \(L\to\infty\). Together, these results reveal a clear separation between transport and memory: angular-momentum transport remains only weakly affected by confinement, whereas torsional memory depends strongly on the radial extent available for stress storage.

The origin of this decoupling becomes even more apparent in the transient dynamics. Figure~\ref{fig_Confinement}d shows the normalized angular velocity, \((\omega-\omega_{\mathrm{ss}})/(\omega(0)-\omega_{\mathrm{ss}})\), on a log--log scale. The curves are only similar at early times, whereas confinement-dependent deviations become pronounced in the subsequent long-time tail. These deviations mark the cutoff of the longest relaxation modes, but in the velocity response they appear only as a slowly decaying contribution. In the recoil, by contrast, the same long-time contributions are integrated over time and therefore strongly affect the accumulated angular displacement after torque removal. Orientational recoil therefore serves as a particularly sensitive readout of the fluid's emergent geometric memory.

\section*{Discussion}
Our results show that the geometry of mechanical driving can create relaxation spectra that are not solely intrinsic material properties. In a viscoelastic fluid with one dominant relaxation time, rotational driving converts radial distance from the probe into a hierarchy of modes. The observed spectrum therefore reflects both local stress relaxation and the spatial extent over which torsional stress can be stored. As the driving time increases, more distant regions of the fluid contribute to the response, in analogy with long-time tails in hydrodynamics \cite{paul1981observation, cichocki2000long}, whereas confinement cuts off this spatial hierarchy and selectively suppresses the longest-lived modes. This interpretation also explains why previous orientational recoil experiments performed at shorter driving times resolved only a single relaxation mode~\cite{wilking2008optically}.

More broadly, our work shows that complex memory need not originate from complex microstructure or disordered energy landscapes alone. It can also emerge when spatial transport and local stress storage are combined with a driving geometry that preserves the organization of deformation: the field is steady in an Eulerian description, while material regions remain registered with it rather than being advected through it. Related forms of geometry-induced memory may therefore arise in systems where localized forcing, transport and slow relaxation remain correlated over space and time, including colloidal glasses~\cite{habdas2025stirring}, near-critical binary mixtures~\cite{hertlein2008direct}, magnetic skyrmion fluids~\cite{pivsljar2022blue}, and cellular matrices~\cite{guo2014probing}. 

The mechanism identified here generalizes mechanically imprinted memory beyond highly organized soft-matter systems, such as liquid-crystal defects and active nematics, where stored stresses and elastic distortions have already been discussed in the context of information storage, logic operations and active microfluidic functions~\cite{kos2022nematic,woodhouse2017active,decamp2015orientational}. Yet, rather than making use of pre-existing orientational order, topological defects or autonomous active flows, the present system generates the relevant stress fields through the driving geometry itself. Mechanically encoded histories may thus be accessible across a broader class of driven viscoelastic soft materials, with potential applications in mechanically addressable memory elements, history-dependent microrheological sensors and rotary micromachines. Arrays of rotating dimers could provide a minimal platform for exploring reservoir-like mechanical information processing, in which driving protocols are written into interacting viscoelastic stress fields and later read out through their collective relaxation dynamics.

%The mechanism identified here generalizes mechanically imprinted memory beyond highly organized soft-matter systems, such as liquid-crystal defects and active nematics, where stored stresses and elastic distortions have already been discussed in the context of information storage, logic operations and active microfluidic functions~\cite{kos2022nematic, woodhouse2017active, decamp2015orientational}. Yet, rather than making use of pre-existing orientational order, topological defects or autonomous active flows, the present system generates the relevant stress fields through the driving geometry itself. Mechanically encoded histories may thus be accessible not only in ordered or architected materials, but across a broader class of driven viscoelastic soft materials. Arrays of rotating dimers could then provide a minimal platform for reservoir-like mechanical information processing, in which driving protocols are written into interacting viscoelastic stress fields and later read out through their collective relaxation dynamics.

\section*{Methods}
\textbf{Fabrication of colloidal dimers:} Superparamagnetic colloids (Dynabeads DM450 Tosylactivated) were first coated with SDS. For this purpose, the particles were dispersed in a $4\, \mathrm{mM}$ SDS solution and kept for $36$ h. Afterwards, the SDS solution was removed and the particles were redispersed in deionized water. To form dimers, the particles were dispersed in the viscoelastic fluid inside the sample cell and a magnetic field was applied parallel to the cell surface. Individual monomers assembled due to in-plane dipole-dipole attraction. The magnetic field was maintained for at least two hours, during which the SDS coating promoted irreversible bonding between the monomers.

\textbf{Fabrication of confinements:}
Confining wells were fabricated by photolithography. A layer (thickness $50 \, \mu\mathrm{m}$) of SU-8 3050 photoresist was spin-coated onto a glass substrate at $3000 \, \mathrm{rpm}$ for $35 \,\mathrm{s}$ with a ramp rate of $200 \, \mathrm{rpm}/\mathrm{s}$. This step defined the height of the confining wells. The coated substrate was then soft-baked at $95 \, ^\circ \, \mathrm{C}$ for $12 \, \mathrm{min}$. The baked photoresist layer was exposed to ultraviolet light for $22 \, \mathrm{s}$ at $260 \, \mathrm{W}$ through a photomask defining circular features of different radii. After exposure, the substrate was post-baked for $4 \, \mathrm{min}$ at  $95 \, ^\circ \, \mathrm{C}$, developed in SU-8 MR-DEV $600$ (PGMEA) for $16$ h and rinsed with isopropanol. After development, the patterned photoresist defined an array of cylindrical wells on the glass surface. The structured substrate served as the bottom of the experimental cell, which was then filled with the viscoelastic fluid containing dispersed colloidal particles. After the dimers had formed inside the cell, they were steered to the centre of the confining wells using spatially inhomogeneous and temporally varying magnetic fields generated by movable permanent magnets around the sample cell.

\section*{Data and code availability}
The source data corresponding to the results presented in the main manuscript and the Supplementary Information, and custom-written codes for plotting, are available from the Zenodo data repository \url{https://doi.org/10.5281/zenodo.22744376} \cite{roy2026geometry}.

\section*{Acknowledgements}
The authors thank Christian Maes for fruitful discussions. 
The authors made limited use of Large Language Models (LLMs) to proofread the text and enhance the clarity and flow of the manuscript. 
CB acknowledges financial support by the ERC Adv. Grant BRONEB (101141477). CB and MK also received funding from the Deutsche Forschungsgemeinschaft, SFB 1432 (425217212), project C5. NR is supported by a scholarship of the Alexander von Humboldt foundation.

%\section*{Funding}

\section*{Author contributions}
N.R. and R.S. contributed equally to the project. N.R. performed the experiments and analyzed the data together with C.B.. R.S. performed analytical and numerical  computations and developed the shell model with M.K..
All authors contributed in writing the manuscript.\\

\section*{Competing interests}
The authors declare no competing interests.

% =========================================================
% Supplementary Information
% =========================================================
\clearpage
\onecolumngrid
\setcounter{equation}{0}
\setcounter{figure}{0}
\setcounter{table}{0}
\hypersetup{hidelinks}

% Keep SI typography locally close to the original 12-pt Times article document.
\begingroup
\fontfamily{ptm}\selectfont
\fontsize{12}{14.4}\selectfont
\setlength{\parindent}{15pt}

\begin{center}
{\fontsize{17.28}{20.5}\selectfont\bfseries
Supplementary Information for\\[0.35em]
Geometry-Controlled Relaxation Spectra in Viscoelastic Fluids\par}
\vspace{2.0em}

{\fontsize{12}{14.4}\selectfont
Niloyendu Roy$^{1}$, Rupayan Saha$^{2}$, Debankur Das$^{2}$,
Matthias Kr{\"u}ger$^{2}$, Clemens Bechinger$^{1}$\par}
\vspace{0.8em}

{\fontsize{10.5}{12.5}\selectfont
$^{1}$Fachbereich Physik, Universität Konstanz, 78457 Konstanz, Germany\\
$^{2}$Institut für Theoretische Physik, Georg-August-Universität Göttingen,
37077 Göttingen, Germany\par}
\end{center}

\vspace{1em}

\listsuppcontents
\clearpage

\vspace{1em}

%For a theoretical understanding, we exploit the symmetries in the flow to model the fluid being composed of concentric spherical shells, viscously coupled to each other by only a dashpot. In addition to flow, such a shell contains two internal subunits, viscoelastically coupled by a dashpot and a spring in parallel. Thus, each shell stores and dissipates elastic energy supplied by the torsional driving of the dimer, and afterwards, act as an independent unit of relaxation (eigenmode). 

\begin{figure}[H]
    \centering
	\includegraphics[width=0.5974\linewidth]{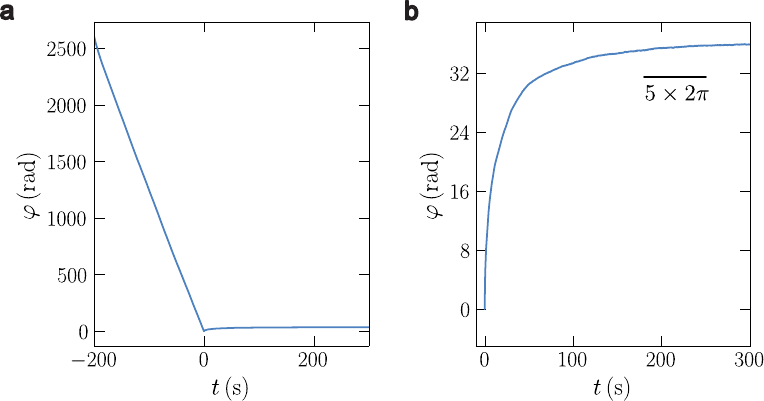}
	\suppcaption{\textbf{S1: Full graph of Fig. 1b of the main manuscript.} \textbf{a} $\varphi$ with respect to time during driving and recoil. \textbf{b} $\varphi$ with respect to time during recoil only ($t>0$). The horizontal bar indicates $\varphi=5\times2\pi \, \mathrm{rad}$, which corresponds to $5$ complete turns of the dimer during recoil.}
    \suppfigentry{S1}{Full graph of Fig. 1b of the main manuscript}
	\label{SI:Fig_S1}
\end{figure}

\suppnote{1}{Comparison between translational and orientational recoil}

\begin{figure}[H]
    \centering
	\includegraphics[width=0.7136\linewidth]{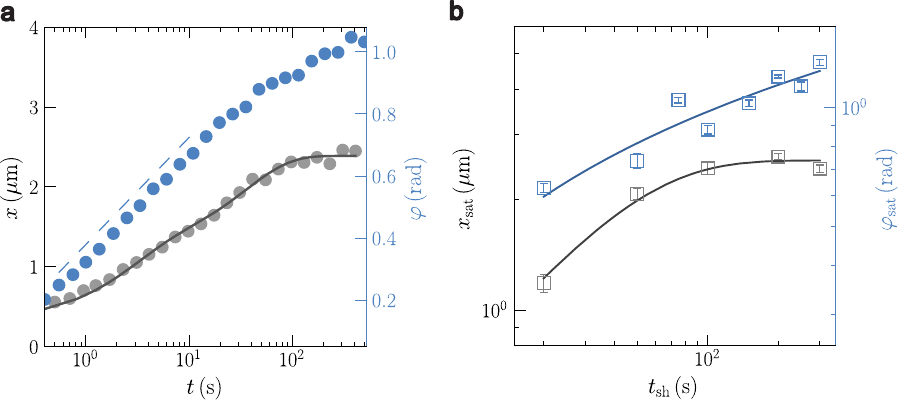}
	\suppcaption{\textbf{S2: Comparison between translational and orientational recoil.} \textbf{a} The grey and the blue circles represent the translational recoil $x (t)$ and the orientational recoil $\varphi (t)$ trajectory of the dimer as a function of time, respectively, for $t_{\textrm{sh}} = 100$ s. Here the dimer's translational coordinate is offset such that $x=0$ when $t=0$. Each recoil trajectory is an ensemble average of $15$ experiments. The grey curve is a fit of the translational recoil trajectory to a double-exponential given by $x (t) = A_s (1 - e^{-t/\tau_s}) + A_l (1 - e^{-t/\tau_l}) $, where $A_s$ and $A_l$ are constants. The dashed line in blue represents, $\varphi(t) = A_1+A_2 \log {t} $, with constants $A_1$ and $A_2$. \textbf{b} The grey and the blue squares represent the  saturation value of the translational ($x_{\mathrm{sat}}$) and orientational recoil ($\varphi_{\mathrm{sat}}$), respectively, as a function of driving time $t_{\textrm{sh}}$. The blue squares represent data points identical to those shown in the inset of Fig. 1c of the main text. The error bars of $x_{\mathrm{sat}}$ correspond to the standard error of mean (SEM) across $7$ experiments. The black solid curve is a fit to the equation $x_{\mathrm{sat}} = B_s(1-e^{-t_{\textrm{sh}}/\tau_s})+B_l(1-e^{-t_{\textrm{sh}}/\tau_l})$, as predicted by a two bath particle model worked out in \suppcite{1}. Here, $B_s$ and $B_l$ are constants. The solid blue curve is a fit to the equation $\varphi _{\mathrm{sat}} = C_1 + C_2 \log t_{\textrm{sh}} $, with constants $C_1$ and $C_2$.}
    \suppfigentry{S2}{Comparison between translational and orientational recoil}
	\label{SI:Fig_S2}
\end{figure}

To establish that torsional forcing produces relaxation dynamics distinct from conventional microrheology, we compare the orientational recoil studied in the main text with the recoil generated by translational forcing. Translational recoil in viscoelastic fluids is a well-established benchmark for probing stress relaxation and has been extensively characterized both experimentally and theoretically. Comparing the two responses therefore provides a direct way to assess whether the long-lived orientational memory observed here reflects a genuinely different relaxation mechanism.

In a translational recoil experiment, a constant force is applied to a colloidal probe for a finite driving time, $-t_{\mathrm{sh}}<t<0$, after which the force is abruptly removed and the subsequent spontaneous motion of the probe is monitored. The recoil originates from the relaxation of viscoelastic stresses accumulated in the surrounding fluid during the driving phase. For the micellar fluid used here, the resulting displacement is known to follow a double-exponential relaxation, characterized by a fast and a slow timescale~\suppcite{2, 1}. These timescales arise from the interplay between the intrinsic structural relaxation of the fluid, which is governed by a single timescale, and the coupling between the fluid and the driven probe. Such double-exponential recoil dynamics have been reported for both spherical colloids~\suppcite{2, 1} and colloidal trimers~\suppcite{3}, indicating that they constitute a signature of translational forcing in this viscoelastic fluid rather than a consequence of probe geometry.

To perform a direct comparison, we carry out translational recoil experiments using the same superparamagnetic dimers employed throughout this work. A constant force is generated by placing a permanent magnet near the sample cell, thereby producing an approximately uniform magnetic-field gradient over the dimensions of the dimer. Because the magnetic field varies only weakly over the length scale of the probe, the dimer experiences an approximately constant force, $\mathbf{F}=(\mathbf{M}\cdot\nabla)\mathbf{H}$. The force is removed by rapidly retracting the magnet from the vicinity of the sample. The protocol is identical to that employed in Ref.~\suppcite{3}.

For the comparison presented here, we choose a driving force $F=2.4 \,\mathrm{pN}$, resulting in a steady-state translational velocity of approximately $0.35 \, \mu \mathrm{m\, s^{-1}}$. This value was selected such that the characteristic shear rates generated by translational driving are comparable to those generated by the torque $\Gamma=7.2\,\mathrm{pN}\mu\mathrm{m}$ used in the orientational recoil experiments, corresponding to a steady-state angular velocity $\omega_{\mathrm{ss}}=0.45 \,\mathrm{rad \, s^{-1}}$ (inset of Fig.~5c of the main text).

The resulting recoil dynamics are shown in Supplementary Fig.~S2a. The translational recoil approaches its asymptotic value through a double-exponential relaxation with characteristic timescales $\tau_s=2.6\,\mathrm{s}$ and $\tau_l=35.2\,\mathrm{s}$, consistent with previous measurements in viscoelastic micellar fluids~\suppcite{4, 1, 3}. These timescales are independent of the driving time $t_{\mathrm{sh}}$ and agree with earlier measurements on colloidal trimers~\suppcite{3}.

The orientational recoil exhibits qualitatively different behaviour. Instead of approaching saturation through a small number of discrete relaxation modes, the recoil grows approximately logarithmically in time and reaches saturation only after hundreds of seconds (Supplementary Fig.~S2a). In a linear--log representation, this behaviour appears nearly as a straight line, in contrast to the pronounced curvature expected for exponential relaxation. The distinction becomes even more apparent when the asymptotic recoil amplitudes are compared as a function of driving time $t_{\mathrm{sh}}$ (Supplementary Fig.~S2b). The translational recoil amplitude $x_{\mathrm{sat}}$ saturates with increasing $t_{\mathrm{sh}}$ and is quantitatively described by a two-bath particle model involving the timescales $\tau_s$ and $\tau_l$~\suppcite{4}. By contrast, the orientational recoil amplitude $\varphi_{\mathrm{sat}}$ continues to increase approximately logarithmically over the experimentally accessible range of $t_{\mathrm{sh}}$, without evidence of an intrinsic cutoff timescale.

These observations demonstrate that the long-lived orientational memory reported in the main text cannot be understood as a simple extension of conventional translational recoil. Whereas translational forcing probes a small number of relaxation modes associated with the material and probe--fluid coupling, torsional forcing generates a broad hierarchy of relaxation times, leading to fundamentally different memory and recoil dynamics.

\suppnote{2}{Shell Model}
\suppsubsection{The  Model}
\suppsubsubsection{Equations of motion}
\begin{figure}[H]
    \centering
	\includegraphics[width=0.95\linewidth]{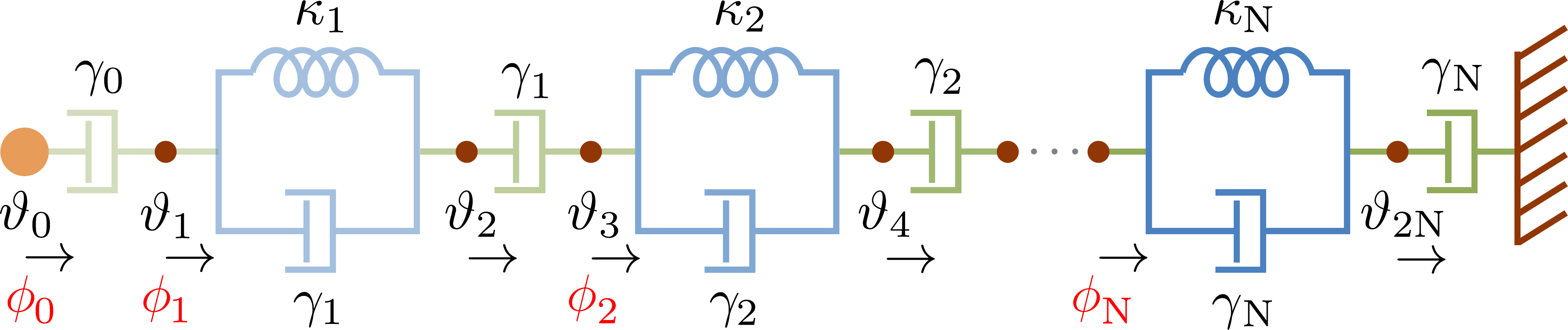}
	\suppcaption{\textbf{S3: \(1\)D representation of the model.} Orange dot denotes the dimer while  brown dots represent \(2N\) nodes. \{\(\vartheta_i\)\} denote the (azimuthal) variables corresponding to the displacements of the nodes. They satisfy the equations of motion in Eq.~\eqref{SI:EQM} which define the model dynamics. Later, we will, for simplicity of illustration, consider the smaller set of variables  \(\{\phi_i\}\) , shown in red,  corresponding to the "shells". The variables $\varphi_i$  mentioned in the main text will finally emerge as  $\varphi_i (t) := \phi_{i} (0) - \phi_{i} (t)$, i.e., the displacements measured after torque is removed. }
    \suppfigentry{S3}{Microscopic Model}
	\label{SI:Fig_SModel}
\end{figure}
The model consists of the dimer connected to a chain of  \(2N\) nodes, via a dashpot \(\gamma_0\). Every pair of nodes is connected by a dashpot. Additionally, a spring is present between every alternating pairs (see Fig.~\textbf{S3}). 
The extension of node $i$ from its ground state is given by $\vartheta_i (t)$, which, in the circular representation,  corresponds to its  angular displacement in the azimuthal direction. The external torque $\Gamma(t)$ acts on the dimer which corresponds to node 0. The equations of motion for  nodes  $0,1,\dots,2N$ are given by, 
%\begin{align} \label{SI:EQM}
%    \begin{split}
%        I_0  \ddot{\vartheta}_0 =& -\gamma_0 (\dot{\vartheta}_{0} - \dot{\vartheta}_{1})   + \Gamma(t),  \\
%        I_i  \ddot{\vartheta}_i =& -\gamma_i (\dot{\vartheta}_{i} - \dot{\vartheta}_{i+1}) -\gamma_{i-1} (\dot{\vartheta}_{i} - \dot{\vartheta}_{i-1}) -\kappa_i (\vartheta_{i} - \vartheta_{i+1}) , \quad  i = 1,3,\dots,2N-1; \\
%        I_i  \ddot{\vartheta}_i =& -\gamma_i (\dot{\vartheta}_{i} - \dot{\vartheta}_{i+1}) -\gamma_{i-1} (\dot{\vartheta}_{i} - \dot{\vartheta}_{i-1}) -\kappa_{i-1} (\vartheta_{i} - \vartheta_{i-1}), \quad  i = 2,4,\dots,2N-2; \\
%        I_{2N}  \ddot{\vartheta}_{2N} =& -\gamma_{2N} \dot{\vartheta}_{2N} -\gamma_{2N-1} (\dot{\vartheta}_{2N} - \dot{\vartheta}_{2N-1}) -\kappa_{2N-1} (\vartheta_{2N} - \vartheta_{2N-1}). 
%    \end{split}
%\end{align}
\begin{align} \label{SI:EQM}
\begin{split}
I_0\ddot{\vartheta}_0
={}&
-\gamma_0
\left(
\dot{\vartheta}_0-\dot{\vartheta}_1
\right)
+\Gamma(t),
\\
I_{2i-1}\ddot{\vartheta}_{2i-1}
={}&
-\gamma_i
\left(
\dot{\vartheta}_{2i-1}-\dot{\vartheta}_{2i}
\right)
-\gamma_{i-1}
\left(
\dot{\vartheta}_{2i-1}-\dot{\vartheta}_{2i-2}
\right)
-\kappa_i
\left(
\vartheta_{2i-1}-\vartheta_{2i}
\right), ~
i=1,\dots,N,
\\
I_{2i}\ddot{\vartheta}_{2i}
={}&
-\gamma_i
\left(
\dot{\vartheta}_{2i}-\dot{\vartheta}_{2i-1}
\right)
-\gamma_i
\left(
\dot{\vartheta}_{2i}-\dot{\vartheta}_{2i+1}
\right)
-\kappa_i
\left(
\vartheta_{2i}-\vartheta_{2i-1}
\right),
~
i=1,\dots,N-1,
\\
I_{2N}\ddot{\vartheta}_{2N}
={}&
-\gamma_N\dot{\vartheta}_{2N}
-\gamma_N
\left(
\dot{\vartheta}_{2N}
-\dot{\vartheta}_{2N-1}
\right)
-\kappa_N
\left(
\vartheta_{2N}
-\vartheta_{2N-1}
\right).
\end{split}
\end{align}
The last node, node $2N$, is connected to a fixed node $2N+1$  with \(\vartheta_{2N+1} = 0\). $I_i$ is the moment of inertia of node $i$; 
%$\gamma_i$ and $\kappa_i$ are the strength of the dashpot and the stiffness of the spring between nodes $i$ and $i+1$, respectively. 
\(\kappa_i\) denotes the stiffness of the spring connecting nodes \(2i-1\) and \(2i\), while \(\gamma_i\) denotes the friction coefficient of both the dashpot connecting these two nodes, i.e., nodes \(2i-1\) and \(2i\), and the dashpot connecting node \(2i\) to node \(2i+1\).
In the relevant overdamped limit, we let  ${\{I_i\} \to 0}$, and  Eq.~(\ref{SI:EQM}) is expressed compactly in matrix-form, 
\begin{align} \label{SI:EQM_Matrix}
    \mathrm{Y} \dot{\bm{\bm{\vartheta}}} + \mathrm{K} \bm{\vartheta} = \bm{\Gamma},
\end{align}
where, the vectors $\bm{\vartheta}, \bm{\Gamma} \in \mathbb{R}^{2N+1}$: $\bm{\vartheta} = (\vartheta_0, \vartheta_1, \dots, \vartheta_{2N})^\top$, $\bm{\Gamma} = (\Gamma(t), 0, \dots, 0)^\top$. The matrices $\mathrm{Y}$ and $\mathrm{K}$ are real and symmetric, representing friction and stiffness respectively. Due to the connectivity, \(\mathrm{Y}\) is tridiagonal and positive definite, whereas \(\mathrm{K}\) is only block-diagonal and positive semi-definite \suppcite{5}. The formal solution of Eq.~\eqref{SI:EQM_Matrix}, with the initial condition $\bm{\vartheta}(t_0)$ at  initial time $t_0$ is given by,
\begin{align} \label{SI:solution_formal}
    \bm{\vartheta}(t) = \exp{(-\mathrm{W} (t-t_0))} \bm{\vartheta}(t_0) + \int_{t_0}^{t}dt^{\prime} \exp{(-\mathrm{W}(t-t^{\prime}))} \mathrm{Y}^{-1} \bm{\Gamma} (t^{\prime}).
\end{align}
Following the analysis presented in Ref.~\suppcite{5}, we identify the normal modes of $\mathrm{W} = \mathrm{Y}^{-1}  \mathrm{K}$ to be the relative displacements between two neighbouring nodes \(\{\delta \vartheta_i\}\),
\begin{align}
    \delta \vartheta_i = \vartheta_i - \vartheta_{i+1}; \implies \vartheta_i = \sum_{j=i}^{2N} \delta \vartheta_j.
\end{align}
Using the normal modes, the angular displacement of any node $i$ to a time-dependent perturbation $\Gamma(t)$ acting on the dimer can be expressed as,
\begin{align} \label{SI:linear_response}
    \vartheta_i (t) = \int_{-\infty}^t dt^{\prime} \chi_{i} (t-t^{\prime}) \Gamma(t^{\prime}); 
\end{align}
where, the susceptibility \(\chi_{i}\), characterizing the causal response at node $i$ due to a perturbation acting on the dimer, is given by,
\begin{align} \label{SI:rf_i}
    \chi_{i} (t-t^{\prime}) = \dfrac{\delta \vartheta_i (t)}{\delta \Gamma(t^{\prime})}.
\end{align}
Depending on the node being even or odd, \(\chi\) has the following form,
\begin{align} 
\chi_{2i-1}(t)
&=
\Theta(t)
\left [
\sum_{j=i}^{N}
\frac{1}{\gamma_j}
+
\sum_{j=i}^{N}
\frac{1}{\gamma_j}
e^{-\kappa_j t/\gamma_j}
\right], \label{SI:rf_odd}\\
\
\chi_{2i}(t)
&=
\Theta(t)
\left[
\sum_{j=i}^{N}
\frac{1}{\gamma_j}
+
\sum_{j=i+1}^{N}
\frac{1}{\gamma_j}
e^{-\kappa_j t/\gamma_j}
\right]. \label{SI:rf_even}
\end{align}
%\begin{align} \label{SI:rf_i}
%    \chi_{i}(t) = \Theta(t) \left(
%    \sum_{\substack{ j \ge i \\ j ~ \text{even} }} \frac{1}{\gamma_j}
%    + \sum_{\substack{j \ge i \\ j ~ \text{odd} }} \dfrac{1}{\gamma_j}e^{-\frac{\kappa_j}{\gamma_j} t}
%    \right).
%\end{align}
%\(\chi_i (t) := {\exp{(-\mathrm{Y}^{-1}  \mathrm{K} t)} \mathrm{Y}^{-1}}_{(i, 0)}\). 
%In particular, we are interested in the response of the dimer under different torque protocols. Hence, we explicitly provide the susceptibility \(\chi_0(t)\),
%\begin{align} \label{SI:rf_dimer_disc}
%    \chi_{0}(t) =  \Theta(t) \left ( \sum_{i=0}^{N/2} \dfrac{1}{\gamma_{2i}} + \sum_{i=1}^{N/2} \dfrac{e^{-(\kappa_{2i-1} / \gamma_{2i-1})t}}{\gamma_{2i-1}} \right).
%\end{align}
We now discuss the behaviour of the odd nodes, i.e., the shell variables. 
\suppsubsubsection{Shell Variables}
While the even nodes are needed for the solution of the model, the discussion of the solution is more intuitive when regarding only the odd nodes, which we refer to as the shell variables, see Fig.~\textbf{S3}. We thus define
%A shell \(i\) is defined to be composed of two adjacent nodes \(2i-1\) and \(2i\) attached by a spring \(\kappa_{i}\) and dashpot \(\gamma_{i}\) in parallel. Two neighboring shells \(i\) and \(i+1\) are thus connected only by a dashpot \(\gamma_{i}\). We further eliminate the even node \(\vartheta_{2i}\) as an internal degree of freedom and relabel the angular displacement of the odd node \(\vartheta_{2i-1}\) as the angular displacement \(\phi_i\) associated with shell \(i\), i.e., 
\[\phi_i \equiv \vartheta_{2i-1}, ~  i = 1, \dots, N.\] 
%Now, we want all the parameters to depend on the shell index \(i\) for which we adopt: \(\kappa_{2i-1} \equiv \kappa_i, ~ \gamma_{2i-1} = \gamma_{2i} \equiv \gamma_{i}; ~  i = 1, \dots, N\). We finally eliminate the intermediate coordinate \(\vartheta_{2i-1}\), associated with shell \(i\), as \(\vartheta_i\). Following Eq.~\eqref{SI:rf_i},  
The susceptibility for  shell \(i\) is thus given by Eq.~\eqref{SI:rf_odd},
%With this input, \(\chi_i\) in 
\begin{align}
    \chi_i(t) 
    = 
    \dfrac{\delta \phi_i (t)}{\delta \Gamma(t^{\prime})} 
    =
    \Theta(t) \left [ \sum_{j=i}^{N}\frac{1}{\gamma_j} (1 + e^{-\kappa_j t/\gamma_j}) \right ]. \label{SI:rf_unit_i}
\end{align}
The dimer's coordinate is relabeled to \(\phi_0\equiv\vartheta_0\). Its susceptibility denoted by \(\chi (t)\), is given by,
\begin{align}
\chi(t)= \Theta(t) \left [\frac{1}{\gamma_0}+ \sum_{i=1}^{N}\frac{1}{\gamma_i} (1 + e^{-\kappa_i t/\gamma_i}) \right ]. \label{SI:rf_dimer}
\end{align}
This is the response function appearing in Eq.~(3) of the main text.

\suppsubsubsection{Recoil Angle}
To appropriately characterize the recoil after a step excitation through torque \(\Gamma(t)\), % i = 0,1, \dots, N
\begin{align} \label{SI:single_step_protocol}
    \Gamma(t) &= \begin{cases}
                    \Gamma_0, \quad  - t_{\text{sh}} < t \le 0 ; \\[4pt]
                    0, \quad t > 0 ; \\
                \end{cases} 
\end{align}
we define the recoil angle \{\(\varphi_i (t) \)\} for shell \(i\), 
\begin{align} \label{SI:recoil_defn}
    \varphi_i (t) \equiv \phi_{i} (0) - \phi_{i} (t). %\quad \delta \varphi_i (t) = \varphi_{i} (t) - \varphi_{i+1} (t). 
\end{align}
To evaluate \(\varphi_i (t) \), we compute \(\phi_i(t)\) using Eq.~\eqref{SI:linear_response}, putting into it the torque-protocol in Eq.~\eqref{SI:single_step_protocol} and the susceptibility in Eq.~\eqref{SI:rf_unit_i}, and subsequently, subtract it from \(\phi_i(0)\). Thus, we have for \(t\geq0\),  
\begin{align} \label{SI:recoil_discrete_shell}
\varphi_i(t)
=
\Gamma_0
\sum_{j=i}^{N}
\frac{1}{\kappa_j}
\left(
1-e^{-\kappa_j t_{\mathrm{sh}}/\gamma_j}
\right)
\left(
1-e^{-\kappa_j t/\gamma_j}
\right).
\end{align}
It is useful to define, as in the main text, the local recoil contribution at shell \(j\),
\begin{align} \label{SI:recoil_local_shell}
\delta\varphi_j(t)
\equiv\varphi_{j} (t) - \varphi_{j+1} (t)=
\frac{\Gamma_0}{\kappa_j}
\left(
1-e^{-\kappa_j t_{\mathrm{sh}}/\gamma_j}
\right)
\left(
1-e^{-\kappa_j t/\gamma_j}
\right).
\end{align}
Hence, the recoil angle of shell \(i\) is the sum of all the local recoil contributions \(\delta  \varphi_j\) for \(j \ge i\): \( \varphi_i(t) = \sum_{j=i}^{N} \delta  \varphi_j\). 
Following the definition in Eq.~\eqref{SI:recoil_defn}, the dimer's recoil at \(t \ge 0\) is denoted by, 
\(\varphi (t) \equiv \phi_{0} (0) - \phi_{0} (t) \).
The same analysis as presented in Eqs.~\eqref{SI:recoil_discrete_shell},~\eqref{SI:recoil_local_shell} results in, 
\begin{align}
\varphi (t) = \sum_{i=1}^{N} \delta  \varphi_i 
= \Gamma_0 \sum_{i=1}^{N} \frac{1}{\kappa_i}
\left(
1-e^{-\kappa_i t_{\mathrm{sh}}/\gamma_i}
\right)
\left(
1-e^{-\kappa_i t/\gamma_i}
\right). 
\end{align}
All local recoil contributions thus add up to produce the total recoil angle of the dimer.

\suppsubsubsection{Continuum Limit}
In the discrete scenario, a shell $i$ has radius $r_i$. $\Delta r$ is the separation between two consecutive shells, and $L = R + N\Delta r$ represents the system-size, with $R$ being the size of the dimer. The continuum limit is taken by considering $N \to \infty$ and $\Delta r \to 0$, such that their product $L-R$  remains finite.  In this limit, the parameters $\gamma_i$ and \(\kappa_i\) respectively become continuous functions \(\gamma(r)\) and  \(\kappa(r)\) of the radial distance $r$ from the origin, i.e., the centre of the dimer. The boundary condition is \(\phi(L) = 0\).

%has an internal structure with a spring $\kappa_i$, and a dashpot $\gamma_i$, to account for the viscoelastic response. Moreover, each unit is connected to its neighbours with a dashpot, e.g., $\gamma_i$ connects units $i$ and $i+1$. \sout{Essentially, we consider \(\gamma_0 \to \infty\) to impose no-slip at the boundary between the dimer and the fluid in contact with it.} This procedure leads to an open boundary condition at $r=R$,  and a Dirichlet boundary condition at $r=L$. 
%Subsequently we will refer to a shell-unit in the continuum limit simply as a shell.
The susceptibility \(\chi\) for a shell at position \(r\) takes the following form in continuum,
\begin{align}
\chi(r,t)
=
\Theta(t) \int_{r}^{L}
\frac{dr^{\prime}}{\gamma(r^{\prime})}
\left[
1+e^{-t/\tau(r^{\prime})}
\right],
\label{SI:rf_shell_cont}
\end{align}
Let \(\tau(r) = \gamma(r)/\kappa(r)\), be the relaxation time at radius \(r\).
Interestingly, the response of a shell at position \(r\) depends on all the shells with position \(r^\prime \ge r \).
Eq.~\eqref{SI:linear_response} can now be expressed with \(\phi_i(t)\) being promoted to a (angular) deformation field \(\phi(r,t)\),
\begin{align} 
    \phi (r, t) = \int_{-\infty}^t dt^{\prime} \chi_{} (r,t-t^{\prime}) \Gamma(t^{\prime});  \label{SI:shell_disp_field}
\end{align}
Next we consider two distinct scenarios: steady-driving and relaxation.
%in Eq.~\eqref{SI:rf_dimer}  essentially involves the fluid confined between two shells of radius, $R$, and $L$ (cf. Fig. ~\ref{fig_ith}) 

\suppsubsection{Steady-driving and scaling of $\gamma(r)$}
First we apply a steady-torque \(\Gamma_0\) at \(t=-t_{\rm sh}\) with \(t_{\rm sh} > 0\), i.e., \(\Gamma (t) = \Gamma_0 \Theta (t+t_{\rm sh})\). Initially, \(\phi(r,-t_{\rm sh})=0\). The angular deformation \(\phi\) and velocity \(\omega = \dot{\phi}\) is given by Eq.~\eqref{SI:shell_disp_field},
\begin{align}
    \phi(r,t) &= \Gamma_0 \int_{r}^{L} dr^\prime \left[ \frac{t+t_{\rm sh}}{\gamma(r^\prime)}  + \frac{1}{\kappa(r^\prime)} \left ( 1 - e^{-\frac{(t+t_{\rm sh})}{\tau(r^\prime)}} \right) \right], \label{SI:shell_phi_cont} \\
    \omega(r,t) &= \Gamma_0 \int_{r}^{L} \frac{dr^\prime}{\gamma(r^\prime)} \left[ 1+e^{-\frac{(t+t_{\rm sh})}{\tau(r^\prime)}} \right] \label{SI:shell_omega_cont}.
\end{align}
%\(\Gamma (t) = \Gamma_0 \Theta (t)\) on the dimer \textcolor{red}{is it not better to stick to the nomenclature of the main text? Switch on torque at $t=-t_{sh}$.},  and 
Steady state is reached at \( (t+t_{\rm sh}) \to \infty \), such that  the second term in the integrand above goes to zero, indicating that steady driving probes solely the viscous coupling \(\gamma(r)\) in the fluid. The steady-state velocity \(\omega(r)\) is given by,
\begin{align}
    \omega(r) = \Gamma_0 \int_{r}^{L} \frac{dr}{\gamma(r)}, \implies \dfrac{d \omega}{dr} = - \dfrac{\Gamma_0}{\gamma (r)}, \label{SI:shell_omega_steady}
\end{align}
Thus, Eq.~\eqref{SI:shell_omega_steady} relates the steady state azimuthal velocity profile in the fluid \(\omega(r)\), which is directly measurable experimentally,  to the dashpot function \(\gamma(r)\). In the following, we derive the dependence of $\gamma(r)$ on $r$ based on simple geometric reasoning.
%For an  in dimensions  \suppcite{6}. %Putting this in Eq.~\eqref{SI:shell_omega_steady}, results n . , and show that it  yields the same scaling.

Consider a spherical shell of radius \(r\) with angular velocity \(\omega(r)\). 
Internal friction generates a torque $M$ opposing the relative rotational motion between adjacent shells,  
\[M (r) =  -\gamma(r) \dfrac{d\omega}{dr} .\]
On the other hand, the frictional torque between adjacent fluid layers is proportional to i) their area of contact, \(S(r)\), ii) to their relative local  tangential velocity, which scales as \(r\,(d\omega/dr)\) and iii) to $r$, relating between local tangential forces and torque. Thus,
\begin{align}
   M(r)
    \propto
    -S(r)\,r^2\,\frac{d\omega}{dr}.
\end{align}
%First we note how the frictional force between two adjacent layers depend: it is proportional to the area of contact \(S(r)\) and the relative velocity \( \dfrac{du}{dr} \) between them,
%\[F_{\mathrm{int}} (r) \propto -S(r) \dfrac{du}{dr}.\] The velocity is related to the angular velocity via \( u \sim r \omega(r) \), and 
The surface area of the spherical shells grow as \( S(r) \propto r^2 \), %the corresponding torque is obtained by multiplying by the force by the radial distance \(r\),  such that,
\begin{align}
    M (r) \propto -r^4 \dfrac{d\omega}{dr}.
\end{align}
Thus, we obtain \(\gamma(r)\sim r^4\) considering spherical shells in dimensions \(d=3\). Plugging this into Eq.~\eqref{SI:shell_omega_steady} yields $\omega(r)\propto r^{-3}$, in agreement with the result of axisymmetric Stokes flow \suppcite{6}, which predicts $\omega(r) \sim r^{-d}$ for \(d = 2, 3.\) Hence we generalize \( \gamma (r) \sim r ^{d+1}\).

\suppsubsection{Relaxation}
We now study the relaxation at \( t \ge 0\) in the fluid after the driving is turned off at \(t=0\). The torque protocol is given in Eq.~\eqref{SI:single_step_protocol}. The recoil angle defined in Eq.~\eqref{SI:recoil_defn} is transformed to  a continuous variable \(\varphi(r,t)\), 
\begin{align}
    \varphi (r,t) := \phi(r,0)-\phi(r,t)
\end{align}
It is given by the continuous analogue of Eq.~\eqref{SI:recoil_discrete_shell}, 
\begin{align} \label{SI:shell_recoil_explicit}
   \varphi(r,t) = \Gamma_0 \int_{r}^{L} \frac{dr^\prime}{\kappa(r^\prime)} \left(1 - e^{-t/\tau(r^\prime)}\right) \left(1 - e^{-t_{\text{sh}}/\tau(r^\prime)}\right) 
\end{align}
Eq.~\eqref{SI:shell_recoil_explicit} reveals that recoil thus exclusively probes the viscoelastic couplings in the fluid.

\suppsubsection{Response of the dimer} 
\suppsubsubsection{Susceptibility}
The scaling relations for \(\gamma(r)\) and \(\kappa(r)\), used for the results given in the main text, are given below,
\begin{align}
    \gamma(r) &= \bar{\gamma} r^{d+1} , \quad d= 3;      \label{SI:gamma_scaing}\\ 
    \kappa (r) &= \bar{\kappa} r ^ \nu , \quad \nu = 1.  \label{SI:kappa_scaing}
\end{align}
The susceptibility of the dimer \(\chi (R,t)\) or simply \(\chi (t)\), as mentioned in Eq.~(4) of the main text, is found by putting \(r =R\) in Eq.~\eqref{SI:rf_shell_cont} and using the explicit scaling relations in Eqs.~\eqref{SI:gamma_scaing},~\eqref{SI:kappa_scaing}, 
%we present below the continuous analogue of Eq.~(3) in the main text,
\begin{align}
    \begin{split}
    \chi (t) &= \int_R^{L} \dfrac{dr}{\gamma(r)} \left[ 1 + e^{-\frac{\kappa(r)}{\gamma (r)} t} \right] \equiv  \int_R^{L} \dfrac{dr}{\bar{\gamma} r^4} \left[ 1 + \exp{ \left (-\frac{\bar{\kappa} t}{\bar{\gamma}r^3} \right ) } \right ] \\
     &= \frac{1}{3 \bar{\gamma}} \left[ \dfrac{1}{R^3}  - \dfrac{1}{L^3} \right] + \frac{1}{3 \bar{\kappa} t} \left[ \exp{\left(-\frac{\bar{\kappa} t}{\bar{\gamma} L^3} \right ) } - \exp{\left(-\frac{\bar{\kappa} t}{\bar{\gamma} R^3} \right ) } \right ]. \label{SI:rf_dimer_L}
\end{split}
\end{align}
%In the following, we take \(d = 3\), and \(\nu = 1\), making \(\alpha = d = 3\). Thus, the susceptibility $\chi(t)$ takes the form,  
%\equiv \frac{1}{3 \bar{\kappa} t} \left( 1 - \exp{(-\frac{t}{\tau_{\min}})} \right);
The first term corresponds to the time-independent instantaneous response \(\chi(t \to \infty)\), governing the steady velocity of the dimer. The second part is the retarded time-dependent response dictating the recoil trajectory. We identify \( \tau_{\min} = {\bar{\gamma} R^3}/{\bar{\kappa}}\), the smallest time-scale (fastest mode) and $\tau_{\max} = {\bar{\gamma} L^3}/{\bar{\kappa}} = \tau_{\min} (\frac{L}{R})^3 $, the longest timescale (slowest mode) coming from the finiteness of the fluid.
For intermediate times, $\tau_{\min} \ll t \ll \tau_{\max}$, the retarded part of the susceptibility \(\chi(t)-\chi(t\rightarrow\infty)\) decays as $\sim 1/t$.

\suppsubsubsection{Velocity} 
The time-dependent angular velocity of the dimer $\omega (R,t)$ or simply \(\omega (t)\) under a steady torque \(\Gamma_0\) turned on at \(t = -t_{\rm sh}\), is simply the susceptibility in Eq.~\eqref{SI:rf_dimer_L} times \(\Gamma_0\) with \(t\) being replaced by \((t+t_{\rm sh})\), the time elapsed after the start of driving, 
\begin{align} \label{SI:dimer_omega_L}
    \omega (t) =  \frac{\Gamma_0}{3 \bar{\gamma}} \left [ \dfrac{1}{R^3}  - \dfrac{1}{L^3} \right ] +   \frac{\Gamma_0}{3 \bar{\kappa} (t+t_{\rm sh})} \left[ \exp{\left(-\frac{\bar{\kappa} }{\bar{\gamma} L^3} (t+t_{\rm sh}) \right )  } - \exp{\left(-\frac{\bar{\kappa}}{\bar{\gamma} R^3} (t+t_{\rm sh}) \right ) }  \right ].
\end{align}
We define the steady state velocity, 
\begin{align} \label{SI:omega_ss}
    \omega_{\rm ss} :=  \lim_{t \to \infty} \omega (t) = \frac{\Gamma_0}{3 \bar{\gamma}} \left [ \dfrac{1}{R^3}  - \dfrac{1}{L^3} \right ].
\end{align} 
The difference \(\omega (t) -  \omega_{\rm ss} \) behaves in the same way as the retarded part of the response and decays as $\sim 1/(t+t_{\rm sh})$ at intermediate times: $\tau_{\min} \ll t+t_{\rm sh} \ll \tau_{\max}$.

%On the contrary, if we somehow confine the dimer in a finite volume of fluid, the power-law decay of the memory kernel  ($\sim t^{-1}$) has a finite cut-off set by the system-size, $L$. Let's refer to the memory kernel of a finite system, $\chi(t;L)$, given by, 
%\begin{align} \label{SI:chi_finite}
%   \chi(t;L) = \frac{1}{3 \bar{\kappa} t} \left[ \exp{(-\frac{t}{\tau_{\max}})} - \exp{(-\frac{t}{\tau_{\min}})} \right].
%\end{align}

\suppsubsubsection{Recoil} 
The key observable in our experiments is the recoil of the dimer \(\varphi(R,t)\). After a step-excitation in Eq.~\eqref{SI:single_step_protocol}, \(\varphi(R,t)\) is given from Eq.~\eqref{SI:shell_recoil_explicit}, 
%\textcolor{red}{arguments confusing, why here $t$ and $t_{sh}$, but above only $t$? Please tell the reader which of these results is used where in the main text. Eq (22), where is it used?}
\begin{align}
     \varphi(R,t) = \Gamma_0 \int_{R}^{L} \frac{dr}{\kappa(r)} (1 - e^{-t/\tau(r)}) (1 - e^{-t_{\text{sh}}/\tau(r)}) \label{SI:dimer_recoil_explicit}
\end{align}
In the main text, we refer to the recoil of the dimer simply as \(\varphi(t)\). The asymptotic recoil amplitude \(\varphi_{\rm sat}\) is defined from Eq.~\eqref{SI:dimer_recoil_explicit} as follows, 
\begin{align}
    \varphi_{\rm sat} := \varphi (t \to \infty) \equiv \lim_{t \to \infty} \varphi(R,t) = \Gamma_0 \int_{R}^{L} \frac{dr}{\kappa(r)} (1 - e^{-t_{\text{sh}}/\tau(r)}).
\end{align}
We calculate the recoil-trajectory of the dimer $ \varphi(t)$ under a step-excitation in Eq.~\eqref{SI:single_step_protocol} with shear-time \(t_{\rm sh}\) in the limit of $L \to \infty$. We make the shear-time explicit in the argument of \(\varphi\) for reasons that will be clear in a moment. \(\varphi(t;t_{\rm sh})\) can be directly obtained from Eq.~\eqref{SI:dimer_recoil_explicit},  
\begin{align}
\varphi(t;t_{\rm sh})
&=
\frac{\Gamma_0}{3\bar{\kappa}}
\Bigg[
\gamma
+\ln\!\left(
\frac{t\,t_{\rm sh}}
{\tau_{\min}(t+t_{\rm sh})}
\right)
-\operatorname{Ei}\!\left(-\frac{t}{\tau_{\min}}\right)
-\operatorname{Ei}\!\left(-\frac{t_{\rm sh}}{\tau_{\min}}\right)
+\operatorname{Ei}\!\left(-\frac{t+t_{\rm sh}}{\tau_{\min}}\right)
\Bigg].
\label{SI:recoil_symmetric}
\end{align}
which explicitly satisfies,
\begin{align}
    \varphi(t;t_{\rm sh})=\varphi(t_{\rm sh};t),   \label{SI:exchange_same} 
\end{align}
reflecting the symmetry under the exchange \(t\leftrightarrow t_{\rm sh}\). 
Eq.~\eqref{SI:recoil_symmetric} may be rewritten as,
\begin{align} \label{SI:recoil_inf}
    \varphi(t; t_{\text{sh}}) =  \dfrac{\Gamma_0 }{3\bar{\kappa}} \left[ \gamma + \ln{ \left( \dfrac{t}{\tau_{\min}} \right) } - \operatorname{Ei}\left(- \dfrac{t}{\tau_{\min}} \right)  + \Delta (t,t_{\text{sh}}) \right],
\end{align}
where, the last term \( \Delta (t,t_{\text{sh}}) \) is  given by,
\[\Delta(t,t_{\text{sh}}) =
\ln\!\left(\frac{t_{\rm sh}}{t+t_{\text{sh}}}\right)
- \operatorname{Ei}\!\left(-\frac{t_{\text{sh}}}{\tau_{\min}}\right)
+ \operatorname{Ei}\!\left(-\frac{t+t_{\text{sh}}}{\tau_{\min}}\right).\]
It is irrelevant in the limit of $t_{\text{sh}} \to \infty$, as \(\lim_{t_{\text{sh}} \to \infty} \Delta(t,t_{\text{sh}}) = 0\). 
In Eq.~\eqref{SI:recoil_inf}, $\gamma$ is the Euler-Mascheroni constant, not to be confused with the dashpot function $\gamma(r)$. The exponential integral $\text{Ei}(x)$ is defined as,
\[ \text{Ei}(x) = \int_{-\infty}^{x} dz \dfrac{e^z}{z}     \]
Finally, the recoil of the dimer for a large shear-time $t_{\text{sh}} \to \infty$ , is given by,
\begin{align}
    \lim_{t_{\text{sh}} \to \infty} \varphi(t; t_{\text{sh}}) =  \dfrac{\Gamma_0 }{3\bar{\kappa}} \left[ \gamma + \ln{(t/\tau_{\min}) - \operatorname{Ei}(-t/\tau_{\min})}  \right]. \label{SI:recoil_ln}
\end{align}
At times \(t \gg \tau_{\min}\), the recoil grows as \(\sim \ln{t}\). 

Invoking the symmetry under the exchange \(t\leftrightarrow t_{\rm sh}\) in Eq.~\eqref{SI:exchange_same}, we may conclude the following, 
\begin{align}
    \lim_{t \to \infty} \varphi(t; t_{\text{sh}}) \equiv \varphi^{\mathrm{sat}}(t_{\text{sh}}) =  \dfrac{\Gamma_0 }{3\bar{\kappa}} \left[ \gamma + \ln{(t_{\text{sh}}/\tau_{\min}) - \operatorname{Ei}(-t_{\text{sh}}/\tau_{\min})}  \right],
\end{align}
which directly follows from Eq.~\eqref{SI:recoil_ln} with \(t\) being replaced by \(t_{\rm sh}\). Hence, for large shear-times \(t_{\rm sh} \gg \tau_{\min}\), the recoil-saturation \( \varphi^{\mathrm{sat}}\) grows as \(\sim \ln{t_{\mathrm{sh}}}\). This explains the observation in the inset of Fig.~1{\bf c} in the main text.

% with the shear time \(t_{\mathrm{sh}}\)
\suppnote{3}{Time-delay in torque reversal}

\begin{figure}[H]
    \centering
	\includegraphics[width=0.6444\linewidth]{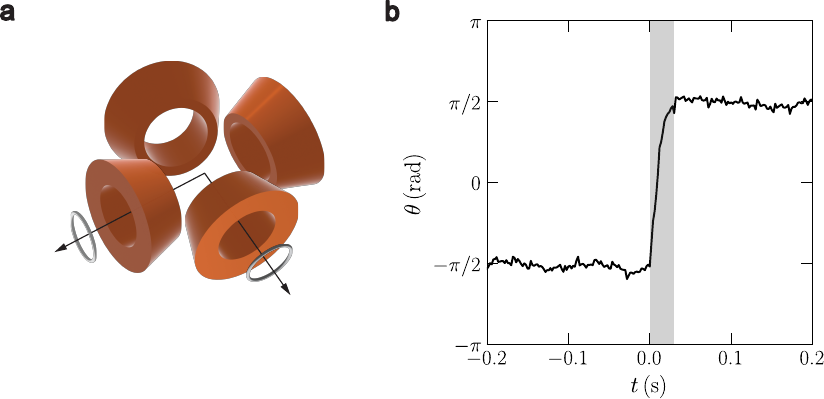}
	\suppcaption{\textbf{S4: Time-delay in torque reversal.} \textbf{a} Schematic diagram of the experimental setup with two pairs of Helmholtz coils, shown in brown, and two pickup coils, shown in grey. \textbf{b} Phase difference $\theta$, between the signals detected at the two pickup coils,  as a function of time. The time axis is chosen such that the switch in the direction of rotation of the magnetic field was intended at $t=0$. The shaded region indicates the time ($\approx33 \, \mathrm{ms}$) required for $\theta$ to reach $\pi/2$.}
    \suppfigentry{S4}{Time-delay in torque reversal}
	\label{SI:Fig_S4}
\end{figure}

Implementing the time-dependent driving protocols requires reversing the direction of the applied torque, which is achieved by reversing the rotation direction of the magnetic field. This is done by switching the phase difference between the two orthogonal magnetic-field components between $\pm \pi/2$. In practice, this switch cannot be instantaneous, because of the finite inductance of the Helmholtz coils used to generate the magnetic field (Fig.~S4a) and the finite response time of the waveform switching performed by the function generator (Tektronix AFG1022). To measure the actual switching time, we monitored the magnetic field with two low-inductance pickup coils placed next to the Helmholtz coils producing the two orthogonal field components (Fig.~S4a). Changes in the magnetic field induce voltages in the pickup coils, which were recorded with a data-acquisition device (NI Instruments).

We then determined the phase difference $\theta$ between the voltages measured by the two pickup coils during a reversal of the magnetic-field rotation. As shown in Fig.~S4b, the switch occurs over a finite interval of approximately $30 \,\mathrm{ms}$. This timescale is much shorter than the characteristic timescales of the dimer dynamics. Therefore, the torque reversals used in the time-dependent driving protocols can be treated as step-like on the timescale relevant to the experiments.

%Implementing the time-dependent driving protocols involves reversing the direction of the torque, caried out by reversing the direction of rotation of the magnetic field. This is achieved by switching the phase difference between the two orthogonal components of the magnetic field between $\pm \pi/2$. This switch however, cannot be instantaneous in reality, and must occur over a finite time, given the finite inductance of the pair of Helmoltz coils (Fig. S3a) that we use to create the magnetic field, as well as the waveform switching mechanism that we use through our function generator (Tektronix, AFG 1022) driving the coils. we measured the real-time change in the magnetic field using two pick-up coils of negligible inductance, held next to the Helmoltz coils (Fig. S3a) that are responsible for two orthogonal components of the magnetic field. A change in the magnetic field causes a change in voltage across the pick-up coils which are measured by a Daq (NI instruments).\\

%We measured the phase difference $\theta$ between the voltages across the two pickup coils, during an intended switch of the direction of rotation of the magnetic field. Fig S3b shows that this switch is not instantaneous, but occurs over a window of $\approx30 \, \mathrm{ms}$. This timescale is however, significantly shorter than the timescales relevant to the dimer's motion. Therefore, the torque reversals during the time-dependent driving protocols, can be considered step-wise.\\

\suppnote{4}{Stress distribution and velocity profile in the fluid under torque reversal}
\suppsubsection{Recoil trajectory and torsional stress under torque reversal}
To compute the recoil trajectory of the dimer $\varphi(t)$ under a multi-step protocol $\Gamma(t)$, defined in the top panels of Fig.~3 in the main text, we first evaluate the azimuthal angle \(\phi(t) \equiv \phi(R,t)\) by putting the multistep form of $\Gamma(t)$ and the susceptibility \(\chi(t)\) of the dimer, as in Eq.~\eqref{SI:rf_dimer_L} taking \(L \to \infty\), in the equation below, 
\[ \phi(t) = \int_{-\infty}^t dt^{\prime} \chi_{} (t-t^{\prime}) \Gamma(t^{\prime}). \]
%is derived by putting the multistep form of $\Gamma(t)$ in Eq.~\eqref{SI:shell_disp_field}, setting \(r=R\) and taking \(L \to \infty\),
Finally the recoil angle of the dimer is simply given by, 
\begin{align} \label{SI:recoil_protocol_dependent}
      \varphi (t) =  \phi(0) - \phi(t) = \int_{R}^{\infty} dr \Sigma (r) (1 - e^{-t/\tau(r)}), 
\end{align}
%the dots in curly braces ${\{ \dots \}}$ denote the protocol under consideration, and 
here, $\Sigma (r)$ is the continuum version of $\delta \varphi_i(t=0)$ (cf. Eq.~\eqref{SI:recoil_local_shell}) under a multi-step protocol, which we take as a measure for the  protocol-dependent stress-distribution in the fluid at $t=0$, i.e., when the torque is removed.  $\Sigma (r)$ is obtained by writing the solution of $\varphi(t)$ in the form of Eq.~\eqref{SI:recoil_protocol_dependent}. This yields, for the different protocols indicated in curly brackets, 
%We explicitly  to obtain the recoil trajectory \(\varphi (R,t)\) of the dimer.\textcolor{red}{ How does the next equations follow from this? where does Eq (35) come from?}

\begin{align}   
    & \Sigma (r)_{\{\Gamma_0,  t_{\rm sh}\}} =  \dfrac{\Gamma_{0}}{\kappa(r)} (1 - e^{-{ t_{\rm sh}}/{\tau (r)}}) \label{SI:recoil_monotonic} \\
    &\Sigma (r)_{\{ \Gamma_{0},  t_{1} ; -\Gamma_0, t_{2} \}} = - \dfrac{\Gamma_{0}}{\kappa(r)} \left[ (1 - e^{-t_{2}/{\tau (r)}}) - e^{-{ t_{2}}/{\tau (r)}} (1 - e^{-{ t_{1}}/{\tau (r)}}) \right ] \label{SI:recoil_nonmonotonic1} \\
    &\Sigma (r)_{\{ \Gamma_{0}, t_{1} ; -\Gamma_0, t_{2}; \Gamma_{0}, t_{3} \}} = \dfrac{\Gamma_{0}}{\kappa(r)} \left[ (1 - e^{-\frac{ t_{3}}{\tau (r)}}) -  e^{-\frac{ t_{3}}{\tau (r)}} (1 - e^{-\frac{t_{2}}{\tau (r)}}) + e^{-\frac{(t_{3}+t_{2})}{\tau (r)}} (1 - e^{-\frac{t_{1}}{\tau (r)}}) \right ] \label{SI:recoil_nonmonotonic2} \\
    \begin{split}
    &\Sigma (r)_{\{ \Gamma_{0}, t_{1} ; -\Gamma_{0}, t_{2}; \Gamma_{0}, t_{3} ; -\Gamma_{0}, t_{4} \}} = \\
    &-\dfrac{\Gamma_{0}}{\kappa(r)} \left[ (1 - e^{-\frac{ t_{4}}{\tau (r)}}) - e^{-\frac{ t_{4}}{\tau (r)}} (1 - e^{-\frac{ t_{3}}{\tau (r)}}) +  e^{-\frac{(t_{4} + t_{3})}{\tau (r)}} (1 - e^{-\frac{t_{2}}{\tau (r)}}) -  e^{-\frac{(t_{4}+t_{3}+t_{2})}{\tau (r)}} (1 - e^{-\frac{t_{1}}{\tau (r)}}) \right ]
    \end{split} \label{SI:recoil_nonmonotonic3}
\end{align}
Here, Eq.~\eqref{SI:recoil_monotonic} reproduces the known result under a single-step excitation (cf. Eq.~\eqref{SI:dimer_recoil_explicit}).

\suppsubsection{Velocity profiles before and after torque reversal}
In Fig.~4 of the main text, The time-dependent torque protocol acting on the dimer is: \(\{\Gamma_0,t_1; -\Gamma_0,t_2\}\), with \(t_{\rm sh} = t_1+t_2\) (see the inset in Fig~4\textbf{b}).
The time-dependent velocity profile before torque reversal is obtained from Eq.~\eqref{SI:shell_omega_cont} with \(L \to \infty\), 
\begin{align}
    \omega(r,t) = \Gamma_0 \int_{r}^{\infty} \frac{dr^\prime}{\gamma(r^\prime)} \left[ 1+e^{-\frac{(t+t_{\rm sh})}{\tau(r^\prime)}} \right], \quad -t_{\rm sh} \le t \le -t_2. \label{SI:velocity_before_reversal}
\end{align}
On the other hand, the time-dependent velocity profile after the torque reversal is given by, 
\begin{align}
    \omega(r,t) = -\Gamma_0 \int_{r}^{\infty} \frac{dr^\prime}{\gamma(r^\prime)} \left[ 1+ \left ( 2 - e^{-\frac{t_1 }{\tau(r^\prime)}} \right) e^{-\frac{(t+t_{\rm sh})}{\tau(r^\prime)}} \right], \quad -t_2 \le t \le 0. \label{SI:velocity_after_reversal}
\end{align}
Eq.~\eqref{SI:velocity_after_reversal} is obtained from Eq.~\eqref{SI:shell_disp_field}, explicitly putting the torque protocol, and taking the time derivative afterwards.
\suppnote{5}{Fit parameters}
Here we provide the parameters used to produce the theoretical curves in the main manuscript.  
\begin{table} [!ht]
    \centering
    \begin{tabular}{|c|c|c|c|}
        \hline
        Fig. & $\bar{\gamma}$ & $\bar{\kappa}$ & \(\Gamma_0\) \\
        \hline
        $3$ & \(0.0014\) & \(2.53\) & \(1\)  \\
        \hline
         $4$ & \(0.0013\) & \(2.1\) & \(0.01\)  \\
        \hline
        $5$\textbf{b} & \(0.0013\) & \(2.1\) & \(1\)  \\
        \hline
        $5$\textbf{c}, $5$\textbf{d} & \(0.0013\) & \(2.1\) & \(0.55\)  \\
        \hline
    \end{tabular}
    \caption{Parameters \(\bar{\gamma}, \, \bar{\kappa}, \, \Gamma_0 \) used to reproduce the experimental curves in Figs.~3, 4, 5 in the main text.}
    \label{SI:tab1}
\end{table}

For the theoretical fits from the shell-model in Fig.~3 of the main text,  the driving times in the multi-step torque protocols are taken as follows: \\
\textbf{a} \((t_1,t_2) = (20, 0.8) \,\mathrm{s} \), \\
\textbf{b} \((t_1,t_2,t_3) = (82,12,1.2) \,\mathrm{s} \), \\
\textbf{c} \((t_1,t_2,t_3,t_4) = (600,75,10,0.8) \,\mathrm{s} \). 

The driving times used to fit Fig.~3\textbf{b} and Fig.~3\textbf{c} vary slightly from the experimental driving times. This adjustment is done to account for limitations of the minimal linear model and the experimental resolution of long-time relaxation. Finally, the recoil trajectories of the dimer in Fig.~3 are obtained from Eq.~\eqref{SI:recoil_protocol_dependent} with \(R = 7 \, \mu {\rm m}\). 

The stress profiles $\Sigma (r)$ at \(t=0\) at the inset of Fig.~3 are obtained from Eqs.~\eqref{SI:recoil_nonmonotonic1}, \eqref{SI:recoil_nonmonotonic2}, \eqref{SI:recoil_nonmonotonic3} using the aforementioned driving times. To enhance the visibility, the stress profiles have been magnified by a factor of \(100\). 

In Fig~5\textbf{b}, we use Eq.~\eqref{SI:dimer_recoil_explicit} to obtain the recoil curves for various confinement sizes \(L\) with \(R = 7 \, \mu {\rm m}\). The time-dependent velocity profiles in Fig~5\textbf{c} and 5\textbf{d} are obtained from Eq.~\eqref{SI:dimer_omega_L}. The steady-state velocity of the dimer as a function of system-size, as in the inset of Fig~5\textbf{c}, is governed by Eq.~\eqref{SI:omega_ss}. 
%with \(\Gamma_0 = 0.55 \,\mathrm{pN}\,\mu\mathrm{m}\).

\suppnote{6}{Orientational recoil at $\Gamma_0=192 \, \mathrm{pN}\mu\mathrm{m}$}

\begin{figure}[H]
    \centering
	\includegraphics[width=0.95\linewidth]{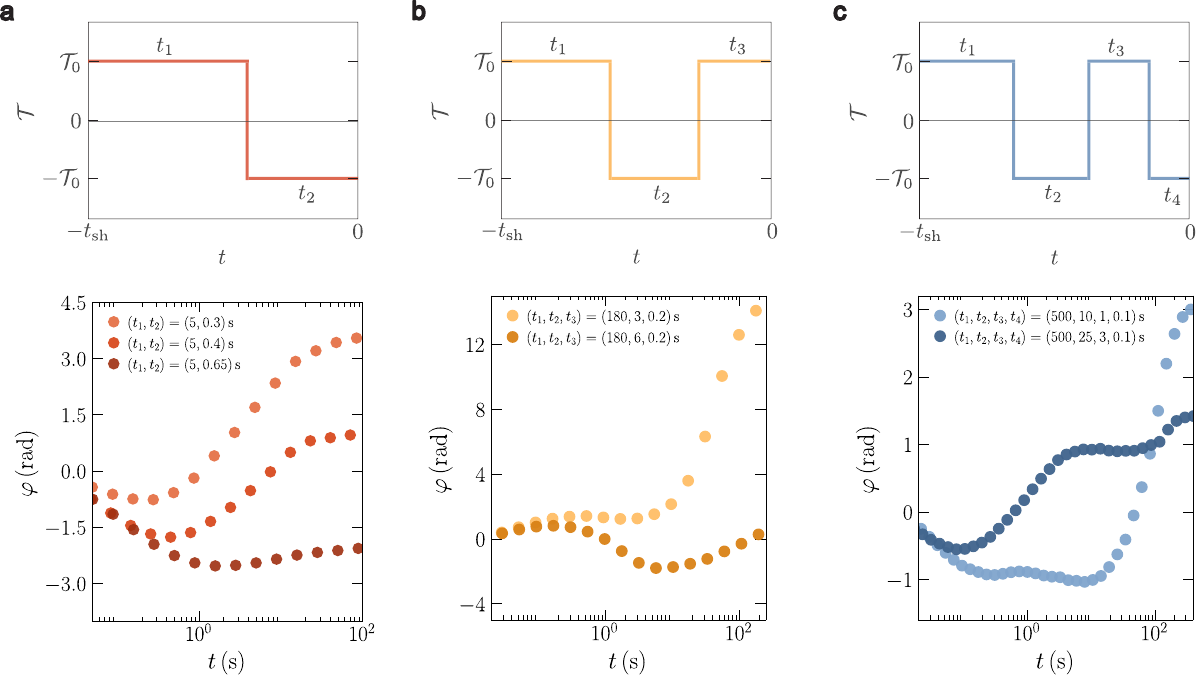}
	\suppcaption{\textbf{S5: Orientational recoils at $\Gamma_0=192\,\mathrm{pN}\mu\mathrm{m}$.} The top panels schematically show the time-dependent torque protocols (time axis not to scale), and the bottom panels show the corresponding recoil responses.}
    \suppfigentry{S5}{Orientational recoils at $\Gamma_0=192\,\mathrm{pN}\mu\mathrm{m}$}
	\label{SI:Fig_S5}
\end{figure}

Non-monotonic recoils after time-dependent torque protocols, as shown in Fig.~3 of the main manuscript, are also observed for the larger torque $\Gamma_0=192 \,\mathrm{pN}\mu\mathrm{m}$, for which steady driving produces large orientational recoils (Fig.~1 of the main manuscript). As shown in Fig.~S5, these non-monotonic responses can involve several complete rotations of the dimer during individual branches of the recoil. Furthermore, systematic changes in the driving protocol lead to corresponding changes in the recoil trajectories, demonstrating that the subsequent recoil encodes information about the preceding torque history.

%Non-monotonic recoils after application of a time-dependent torque protocol, as shown in Fig. 3 of the main manuscript, can also be obtained for $\Gamma_0=-192 \, \mathrm{pN}\mu\mathrm{m}$, which results in large recoil responses of the dimer upon steady driving (Fig. 1 of main manuscript). Fig. S4 shows that the non-monotonic recoils can even occur with the dimer rotating by full turns during one of its branches. Furthermore, a systematic variation of the protocol results into systemic variations of the recoil trajectories, thereby enabling us to use these recoil responses, to predict the driving history.

\suppnote{7}{Orientational recoil away from sample cell surface}

\begin{figure}[H]
    \centering
	\includegraphics[width=0.6556\linewidth]{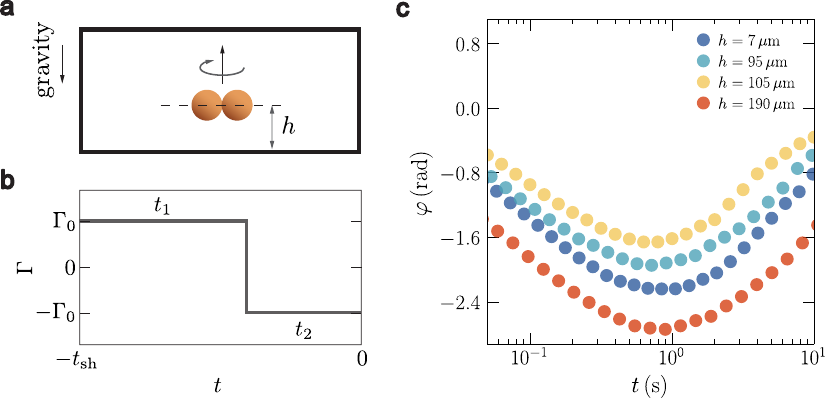}
	\suppcaption{\textbf{S6: Non-monotonic recoil away from surface.}  \textbf{a} Schematic diagram of the dimer sedimenting through the sample cell with thickness $200 \, \mu\mathrm{m}$. \textbf{b} Schematic of the time-dependent driving protocol, with $\Gamma_0 = 192 \, \mathrm{pN}\mu\mathrm{m}$, $ t_1 = 10 \, \mathrm{s}$, and $ t_2 = 0.5 \, \mathrm{s}$. \textbf{c}, $\varphi$ as a function of time during recoil. $h$ (mentioned in the legend) is measured at $t=0$.}
    \suppfigentry{S6}{Non-monotonic recoil away from surface}
	\label{SI:Fig_S6}
\end{figure}

To assess the role of the bottom surface of the sample cell in the orientational recoil of the dimer, we performed non-monotonic recoil experiments at different heights $h$ using a sedimenting dimer (Fig.~S6a). The dimer sedimented at a rate of approximately $0.1\,\mu\mathrm{ms^{-1}}$. We therefore used time-dependent driving protocols (Fig.~S6b) for which the entire driving sequence and the subsequent recoil reversal occurred within a time window of about $20\,\mathrm{s}$. During this time, the dimer height changed by only $\approx 2\,\mu\mathrm{m}$, allowing us to probe the fluid response at an approximately fixed height within the sample cell. Since the rotating magnetic field lies in the horizontal plane, it aligns the dimer within this plane during driving. This alignment is maintained during relaxation, enabling us to track the dimer orientation in the horizontal plane throughout the recoil. Figure~S6c shows that the resulting non-monotonic recoil responses are qualitatively independent of $h$. This indicates that the mechanism underlying the geometry-generated relaxation spectrum is not controlled by the nearby bottom surface, although the main experiments are performed close to it.

\suppnote{8}{Orientational recoil in polymer solution}

\begin{figure}[H]
    \centering
	\includegraphics[width= 0.6204\linewidth]{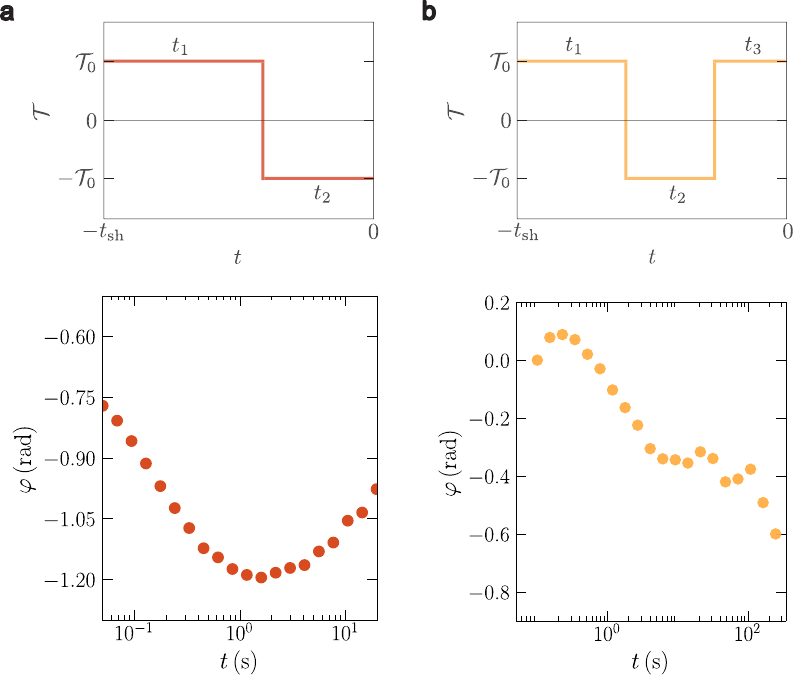}
	\suppcaption{\textbf{S7: Recoils in polymer solution.} The top panels in \textbf{a}-\textbf{b} schematically show the time-dependent torque protocols, and the bottom panels show the corresponding recoil responses of the dimer. In \textbf{a}, $(t_1,t_2)=(20,0.5)\, \mathrm{s}$; in \textbf{b}, $(t_1,t_2,t_3)=(200,1.5,0.1)\, \mathrm{s}$.  }
    \suppfigentry{S7}{ Recoils in polymer solution}
	\label{SI:Fig_S7}
\end{figure}

To test whether the effects of microscale torsion are specific to wormlike micellar solutions, we performed orientational recoil experiments with the dimer in a semidilute solution of polyacrylamide (PAAM) at a concentration of $0.015\,\mathrm{wt}\%$. This solution exhibits pronounced viscoelasticity and has previously been shown to produce double-exponential translational recoil of colloidal probes with timescales $\approx 3 \, \mathrm{s}$ and $60 \, \mathrm{s}$~\suppcite{3}. Figure~S7 shows non-monotonic orientational recoil responses of the dimer, with one or two reversals of the recoil direction following driving protocols with one or two torque reversals, respectively. A recoil trajectory with two direction changes cannot be explained by a fluid response containing only one or two relaxation times~\suppcite{7}. The results in Fig.~S7 therefore support the conclusion that torsional forcing generates a broad relaxation spectrum, and that this effect is not limited to the micellar fluid used in the main experiments.

%To show that the unique effects of microscale torsion  is not limited to the micellar solutions, we perform orientational recoil experiments with the dimer in semi dilute solution of polymer polyacrylamide (PAAM) with concentration $0.015 \, \mathrm{wt}\%$. The solution shows pronounced viscoelasticity and have been characterized to produce double exponential recoil motion of a colloidal probe upon translational \suppcite{3}. Fig. S6 shows non-monotonic recoil responses of the dimer that changes its direction one or twice, in response to torque reversal, approximately once or twice, respectively. Please note that, a recoil response changing its direction twice, cannot be explained by eigenmodes of the fluid with one or two timescales of relaxation \suppcite{vaidya2025observation}. Therefore, the results in Fig. S6b, proves that the torsional forcing generates a broad relaxation spectra.  

\suppnote{9}{Tracer's motion during dimer's recoil}

\begin{figure}[H]
    \centering
	\includegraphics[width= 0.6731
\linewidth]{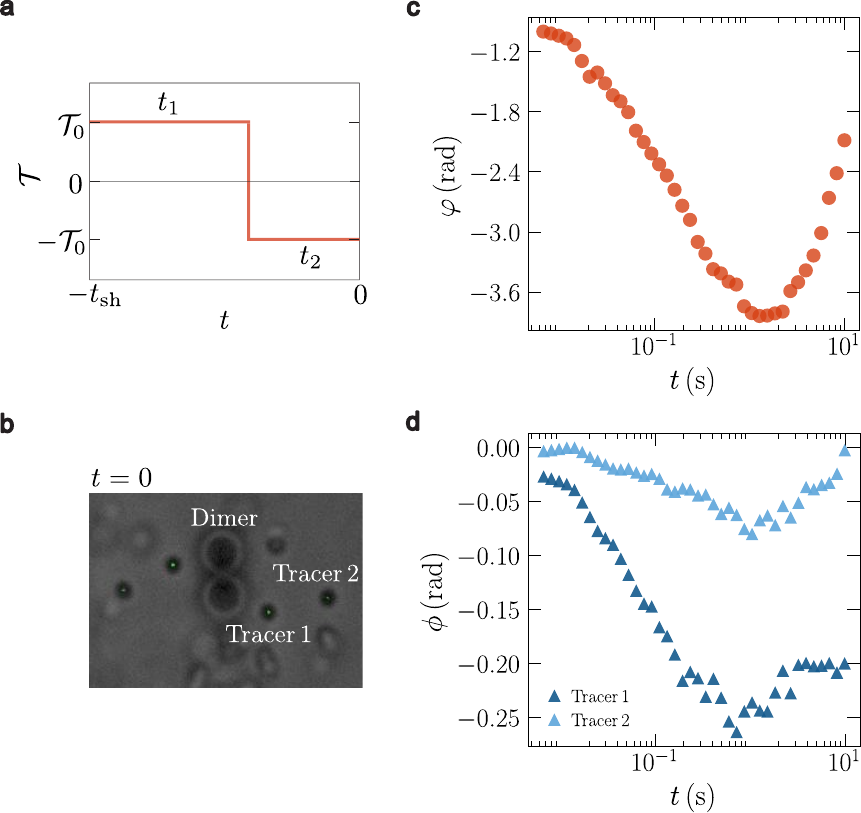}
	\suppcaption{\textbf{S8: Tracer particle motion during fluid's relaxation.} \textbf{a} Schematic representation of the driving protocol applied to the dimer. Here, $\Gamma_0 = 192 \, \mathrm{pN}\mu\mathrm{m}$, $t_1 = 30 \, \mathrm{s}$, and $\mathrm{t_2}=1 \, \mathrm{s}$. \textbf{b} A snapshot of the dimer along with non-magnetic silica tracer particles of diameter $1.8$ $\mu$m, at $t = 0$. The snapshot is from Supplementary Video 9. Here the tracer particles labeled as $\mathrm{Tracer \, 1}$, and $\mathrm{Tracer \, 2}$ are situated at a distance $7$ $\mu$m, and $14$ $\mu$m, respectively,  from the center of mass of the dimer at $t=0$. The green dots represent tracked particles.  \textbf{c} $\varphi$ as a function of time of the dimer. \textbf{d} The azimuthal coordinate of the tracer particles (mentioned in the legend) $\phi$ as a function of time. }
    \suppfigentry{S8}{Tracer particle motion during fluid's relaxation}
	\label{SI:Fig_S8}
\end{figure}

The flow field of tracer particles during dimer driving provides information about the outward transport of angular momentum, whereas tracer motion during dimer recoil probes the relaxation of stored torsional stress. A particularly interesting scenario occurs when the dimer undergoes a non-monotonic recoil in response to a time-dependent driving protocol. The theoretical model predicts that, depending on the driving protocol, relaxation can involve not only the relaxation of the azimuthal stretches $\delta \varphi_i$ of different shells, but also non-monotonic motion of the shells themselves. One such instance is shown in Supplementary Video~7. We find experimental signatures of this behaviour in the non-monotonic motion of tracer particles. To measure trajectories over sufficiently long times, we used silica tracer particles with a diameter of $1.8,\mu\mathrm{m}$, which exhibit lower Brownian diffusion than the $0.7,\mu\mathrm{m}$ tracers used in Fig.~4 of the main manuscript and therefore allow more reliable long-time tracking. The torque protocol applied to the dimer is schematically shown in Fig.~S8a; the corresponding non-monotonic recoil of the dimer is shown in Fig.~S8c. This non-monotonicity is also reflected in the tracer-particle trajectories (Supplementary Video~9); the tracer particles reverse their azimuthal motion at approximately the same time as the dimer reverses its recoil direction, as quantified by their azimuthal coordinate $\phi(t)$ in Fig.~S8d.

%While the flow-field of tracer particles during the driving of the dimer, elucidate on the transport of angular momentum outwards into the fluid, motion of the tracer particles during the dimer's recoil can elucidate the relaxation mechanism of the fluid. One particularly interesting scenario is when the dimer undergoes a non-monotonic recoil in response to a time-dependent driving protocol.  The theoretical model predicts that depending on the driving protocol, there could be relaxation phases, where in addition to the annealing of the azimuthal stretches ($\delta \varphi_i$) of different shells, the shells themselves can undergo non-monotonic motion. One such instance can be seen in Supplementary Video 7. We find experimental signatures of this in non-monotonic motion of the tracer particles. In order to measure trajectories over a long time, we chose silica tracer particles of diameter $1.8$ $\mu$m, that has lesser Brownian diffusion as compared to tracer particles of diameter $0.7$ $\mu$m (as in Fig. 4 of the main manuscript), as well as offers higher tracking resolution. The torque protocol applied to the dimer is schematically represented in Fig. S5a. This results in a non-monotonic recoil of the dimer, as shown in Supplementary Fig. S5c. This non-monotonicity is reminiscent in the tracer particle trajectories (Supplementary Video SV9), that changes its direction of motion at around the same time as the reversal of the dimer's recoil, which is reflected in the azimuthal coordinate of the tracer particles as shown in Fig. S7.

\singlespacing

\newpage
\suppvideosection{Description of additional Supplementary files}

\large{\textbf{Supplementary video 1:}} Dimer's motion during driving with a steady torque $\Gamma_0 = 192 \, \mathrm{pN}\mu\mathrm{m}$, and corresponding orientational recoil. The left panel shows a snippet of the driving, carried out for $t_{\mathrm{sh}}=180 \, \mathrm{s}$. The right panel shows the first four complete turns of the dimer during its orientational recoil. Both the videos are shown in real time. \\
\\
\large{\textbf{Supplementary video 2:}} The shell model is demonstrated during a monotonic driving for $t_{\mathrm{sh}}=5\,\mathrm{s}$, and the following relaxation. The yellow line segment in the middle represents the dimer. Each shell is represented with a dashed circle, color coded according to their index. The stretches of the torsional springs in each shell are shown by the solid arcs between the short yellow line segments. The arcs are color coded according to the corresponding torsional stretch $\delta \varphi_i$. The top panel shows the time evolution of the dimer's orientation. The model parameters chosen here to create the video, are not identical to the ones used for the fitting of the experimental data. These parameters are chosen in order to visually demonstrate the mechanism. \\
\\
\large{\textbf{Supplementary videos 3,4}} The shell model is demonstrated for  time-dependent driving, with one (Supplementary video 3) and two (Supplementary video 4) torque reversals, given by the protocols $(t_1,t_2)=(50,3) \, \mathrm{s}$, and $(t_1,t_2,t_3)=(200,15,1.5) \, \mathrm{s}$, respectively. The color code of $\delta \varphi _i$ facilitates visualizing the spatial distribution of the stress stored into the fluid by the time-dependent driving protocol. The model parameters have been chosen identical to Supplementary video 2. The dimer's orientation is shown in the top panels.\\
\\
\large{\textbf{Supplementary videos 5--7:}} Video of the dimer performing non-monotonic orientational recoil in response to a driving protocol, where the direction of the external torque was reversed one (Supplementary video 5), two (Supplementary video 6) or three (Supplementary video 7) times.  Here $\Gamma_0 = 192 \, \mathrm{pN}\mu\mathrm{m}$. The videos have been rendered with a custom timeline. The red bar at the bottom of the video represents the progression of time. The yellow line segment represents the orientation and the position of the dimer.  The dimer can be seen reversing the direction of its recoil motion corresponding to the number of torque reversals during driving. The direction of the recoil motion is shown by the white arrows. Supplementary video 5 shows that the non-monotonic orientational recoils can as well occur by multiple complete rotations of the dimer.    \\
\\
\large{\textbf{Supplementary video 8}} This video demonstrates the motion of non-magnetic polystyrene tracer particles of diameter $2.7 \, \mu\mathrm{m}$, as the dimer is first rotated in the clockwise direction for $200 \,s $, with $\Gamma_0 = 192 \, pN\mu\mathrm{m}$, and then in the counter-clockwise direction for $200 \, \mathrm{s}$, following an instant reversal in the direction of the torque. The arrows at the upper left denote the direction of torque. The video is shown $6.6$ times faster than the real time. At the moment of torque reversal, all tracer particles, irrespective of the distance from the dimer, for instance the ones even $\approx100 \, \mu\mathrm{m}$ away, reverse their azimuthal velocity $\mathcal{W}$ instantly.\\
\\
\large{\textbf{Supplementary video 9}} This video shows the motion of non-magnetic silica tracer particles of diameter $1.85 \, \mu\mathrm{m}$, during a non-monotonic orientational recoil of the dimer following a driving protocol defined by $(t_1,t_2)=(30,1)\, \mathrm{s}$. Non-monotonic motion of the tracer particles can clearly be observed. The arrow on the upper left denotes the direction of the dimer's orientational recoil at every time instant. The video is rendered with a custom timeline, shown by the label and the progress bar at the bottom. The time evolution of the azimuthal coordinate $\phi$ of the tracer particles, and the orientation of the dimer during the recoil is shown in Supplementary Figure S8.

\clearpage
\suppsectionheading{Supplementary References}

\begingroup
\fontsize{10.5}{12.5}\selectfont
\setlength{\parindent}{0pt}
\newcounter{suppref}
\begin{list}{[\arabic{suppref}]}{%
  \usecounter{suppref}%
  \setlength{\leftmargin}{2.4em}%
  \setlength{\labelwidth}{2.0em}%
  \setlength{\labelsep}{0.4em}%
  \setlength{\itemsep}{0.55em}%
  \setlength{\parsep}{0pt}%
  \setlength{\topsep}{0.4em}%
}

\item Caspers, J. \emph{et al.} How are mobility and friction related in viscoelastic fluids?
\emph{The Journal of Chemical Physics} \textbf{158}, 024901 (2023).
URL \url{https://doi.org/10.1063/5.0129639}.

\item Ginot, F., Caspers, J., Kr{\"u}ger, M. \& Bechinger, C.
Barrier crossing in a viscoelastic bath.
\emph{Physical Review Letters} \textbf{128}, 028001 (2022).

\item Cao, X. \emph{et al.} Memory-induced magnus effect.
\emph{Nature Physics} \textbf{19}, 1904--1909 (2023).

\item Ginot, F. \emph{et al.} Recoil experiments determine the eigenmodes of viscoelastic fluids.
\emph{New Journal of Physics} \textbf{24}, 123013 (2022).
URL \url{https://doi.org/10.1088/1367-2630/aca8c7}.

\item Saha, R. \& Kr{\"u}ger, M. The peculiar response of Kelvin--Voigt chains with a free end.
\emph{Journal of Physics A: Mathematical and Theoretical} \textbf{59}, 315003 (2026).
URL \url{https://doi.org/10.1088/1751-8121/ae8b46}.

\item Dhont, J. K. \emph{An introduction to dynamics of colloids}, vol.~2
(Elsevier, 1996).

\item Vaidya, S. S., Muruga, L., Caspers, J., Kr{\"u}ger, M. \& Bechinger, C.
Observation and control of nonmonotonic recoils in a viscoelastic fluid.
\emph{Phys. Rev. Res.} \textbf{7}, 033084 (2025).
URL \url{https://link.aps.org/doi/10.1103/74nx-3t1h}.

\end{list}
\endgroup

\endgroup % end local Supplementary Times/12-pt typography


\begin{thebibliography}{40}%
\makeatletter
\providecommand \@ifxundefined [1]{%
 \@ifx{#1\undefined}
}%
\providecommand \@ifnum [1]{%
 \ifnum #1\expandafter \@firstoftwo
 \else \expandafter \@secondoftwo
 \fi
}%
\providecommand \@ifx [1]{%
 \ifx #1\expandafter \@firstoftwo
 \else \expandafter \@secondoftwo
 \fi
}%
\providecommand \natexlab [1]{#1}%
\providecommand \enquote  [1]{``#1''}%
\providecommand \bibnamefont  [1]{#1}%
\providecommand \bibfnamefont [1]{#1}%
\providecommand \citenamefont [1]{#1}%
\providecommand \doibase [0]{https://doi.org/}%
\providecommand \selectlanguage [0]{\@gobble}%
\providecommand \bibinfo  [0]{\@secondoftwo}%
\providecommand \bibfield  [0]{\@secondoftwo}%
\providecommand \translation [1]{[#1]}%
\providecommand \BibitemOpen [0]{}%
\providecommand \bibitemStop [0]{}%
\providecommand \bibitemNoStop [0]{.\EOS\space}%
\providecommand \EOS [0]{\spacefactor3000\relax}%
\providecommand \BibitemShut  [1]{\csname bibitem#1\endcsname}%
\let\auto@bib@innerbib\@empty
%</preamble>
\bibitem [{\citenamefont {Ferry}(1980)}]{ferry1980viscoelastic}%
  \BibitemOpen
  \bibfield  {author} {\bibinfo {author} {\bibfnamefont {J.~D.}\ \bibnamefont
  {Ferry}},\ }\emph {\bibinfo {title} {Viscoelastic Properties
  of Polymers}},\ \bibinfo {edition} {3rd}\ ed.\ (\bibinfo  {publisher} {John
  Wiley \& Sons},\ \bibinfo {address} {New York},\ \bibinfo {year}
  {1980})\BibitemShut {NoStop}%
\bibitem [{\citenamefont {Larson}(2005)}]{larson2005rheology}%
  \BibitemOpen
  \bibfield  {author} {\bibinfo {author} {\bibfnamefont {R.~G.}\ \bibnamefont
  {Larson}},\ }\bibfield  {title} {\bibinfo {title} {The rheology of dilute
  solutions of flexible polymers: {P}rogress and problems},\ }\href
  {https://doi.org/10.1122/1.1835336} {\bibfield  {journal} {\bibinfo
  {journal} {Journal of Rheology}\ }\textbf {\bibinfo {volume} {49}},\ \bibinfo
  {pages} {1} (\bibinfo {year} {2005})}\BibitemShut {NoStop}%
\bibitem [{\citenamefont {Chaudhuri}\ \emph {et~al.}(2007)\citenamefont
  {Chaudhuri}, \citenamefont {Parekh},\ and\ \citenamefont
  {Fletcher}}]{chaudhuri2007reversible}%
  \BibitemOpen
  \bibfield  {author} {\bibinfo {author} {\bibfnamefont {O.}~\bibnamefont
  {Chaudhuri}}, \bibinfo {author} {\bibfnamefont {S.~H.}\ \bibnamefont
  {Parekh}},\ and\ \bibinfo {author} {\bibfnamefont {D.~A.}\ \bibnamefont
  {Fletcher}},\ }\bibfield  {title} {\bibinfo {title} {Reversible stress
  softening of actin networks},\ }\href {https://doi.org/10.1038/nature05459}
  {\bibfield  {journal} {\bibinfo  {journal} {Nature}\ }\textbf {\bibinfo
  {volume} {445}},\ \bibinfo {pages} {295} (\bibinfo {year}
  {2007})}\BibitemShut {NoStop}%
\bibitem [{\citenamefont {Zhou}\ and\ \citenamefont
  {Schroeder}(2018)}]{zhou2018dynamically}%
  \BibitemOpen
  \bibfield  {author} {\bibinfo {author} {\bibfnamefont {Y.}~\bibnamefont
  {Zhou}}\ and\ \bibinfo {author} {\bibfnamefont {C.~M.}\ \bibnamefont
  {Schroeder}},\ }\bibfield  {title} {\bibinfo {title} {Dynamically
  heterogeneous relaxation of entangled polymer chains},\ }\href
  {https://doi.org/10.1103/PhysRevLett.120.267801} {\bibfield  {journal}
  {\bibinfo  {journal} {Phys. Rev. Lett.}\ }\textbf {\bibinfo {volume} {120}},\
  \bibinfo {pages} {267801} (\bibinfo {year} {2018})}\BibitemShut {NoStop}%
\bibitem [{\citenamefont {Kovacs}(1963)}]{kovacs1963glass}%
  \BibitemOpen
  \bibfield  {author} {\bibinfo {author} {\bibfnamefont {A.~J.}\ \bibnamefont
  {Kovacs}},\ }\bibfield  {title} {\bibinfo {title} {Glass transition in
  amorphous polymers: {A} phenomenological study},\ }\bibfield
  {journal} {\bibinfo  {journal} {Adv. Polym. Sci.}\ }\textbf {\bibinfo
  {volume} {3}},\ \bibinfo {pages} {394} (\bibinfo {year} {1963})\BibitemShut
  {NoStop}%
\bibitem [{\citenamefont {Mandal}\ \emph {et~al.}(2021)\citenamefont {Mandal},
  \citenamefont {Tapias},\ and\ \citenamefont {Sollich}}]{Mandal2021}%
  \BibitemOpen
  \bibfield  {author} {\bibinfo {author} {\bibfnamefont {R.}~\bibnamefont
  {Mandal}}, \bibinfo {author} {\bibfnamefont {D.}~\bibnamefont {Tapias}},\
  and\ \bibinfo {author} {\bibfnamefont {P.}~\bibnamefont {Sollich}},\
  }\bibfield  {title} {\bibinfo {title} {Memory in non-monotonic stress
  response of an athermal disordered solid},\ }\href
  {https://doi.org/10.1103/PhysRevResearch.3.043153} {\bibfield  {journal}
  {\bibinfo  {journal} {Phys. Rev. Res.}\ }\textbf {\bibinfo {volume} {3}},\
  \bibinfo {pages} {043153} (\bibinfo {year} {2021})}\BibitemShut {NoStop}%
\bibitem [{\citenamefont {Keim}\ \emph {et~al.}(2019)\citenamefont {Keim},
  \citenamefont {Paulsen}, \citenamefont {Zeravcic}, \citenamefont {Sastry},\
  and\ \citenamefont {Nagel}}]{Keim2019}%
  \BibitemOpen
  \bibfield  {author} {\bibinfo {author} {\bibfnamefont {N.~C.}\ \bibnamefont
  {Keim}}, \bibinfo {author} {\bibfnamefont {J.~D.}\ \bibnamefont {Paulsen}},
  \bibinfo {author} {\bibfnamefont {Z.}~\bibnamefont {Zeravcic}}, \bibinfo
  {author} {\bibfnamefont {S.}~\bibnamefont {Sastry}},\ and\ \bibinfo {author}
  {\bibfnamefont {S.~R.}\ \bibnamefont {Nagel}},\ }\bibfield  {title} {\bibinfo
  {title} {Memory formation in matter},\ }\href
  {https://doi.org/10.1103/RevModPhys.91.035002} {\bibfield  {journal}
  {\bibinfo  {journal} {Rev. Mod. Phys.}\ }\textbf {\bibinfo {volume} {91}},\
  \bibinfo {pages} {035002} (\bibinfo {year} {2019})}\BibitemShut {NoStop}%
\bibitem [{\citenamefont {Zia}\ and\ \citenamefont
  {Brady}(2013)}]{zia2013stress}%
  \BibitemOpen
  \bibfield  {author} {\bibinfo {author} {\bibfnamefont {R.~N.}\ \bibnamefont
  {Zia}}\ and\ \bibinfo {author} {\bibfnamefont {J.~F.}\ \bibnamefont
  {Brady}},\ }\bibfield  {title} {\bibinfo {title} {Stress development,
  relaxation, and memory in colloidal dispersions: Transient nonlinear
  microrheology},\ }\href {https://doi.org/10.1122/1.4775349} {\bibfield
  {journal} {\bibinfo  {journal} {Journal of Rheology}\ }\textbf {\bibinfo
  {volume} {57}},\ \bibinfo {pages} {457} (\bibinfo {year} {2013})}\BibitemShut
  {NoStop}%
\bibitem [{\citenamefont {Vaidya}\ \emph {et~al.}(2025)\citenamefont {Vaidya},
  \citenamefont {Muruga}, \citenamefont {Caspers}, \citenamefont {Kr\"uger},\
  and\ \citenamefont {Bechinger}}]{Vaidya2025}%
  \BibitemOpen
  \bibfield  {author} {\bibinfo {author} {\bibfnamefont {S.~S.}\ \bibnamefont
  {Vaidya}}, \bibinfo {author} {\bibfnamefont {L.}~\bibnamefont {Muruga}},
  \bibinfo {author} {\bibfnamefont {J.}~\bibnamefont {Caspers}}, \bibinfo
  {author} {\bibfnamefont {M.}~\bibnamefont {Kr\"uger}},\ and\ \bibinfo
  {author} {\bibfnamefont {C.}~\bibnamefont {Bechinger}},\ }\bibfield  {title}
  {\bibinfo {title} {Observation and control of nonmonotonic recoils in a
  viscoelastic fluid},\ }\href {https://doi.org/10.1103/74nx-3t1h} {\bibfield
  {journal} {\bibinfo  {journal} {Phys. Rev. Res.}\ }\textbf {\bibinfo {volume}
  {7}},\ \bibinfo {pages} {033084} (\bibinfo {year} {2025})}\BibitemShut
  {NoStop}%
\bibitem [{\citenamefont {Creton}\ and\ \citenamefont
  {Ciccotti}(2016)}]{creton2016fracture}%
  \BibitemOpen
  \bibfield  {author} {\bibinfo {author} {\bibfnamefont {C.}~\bibnamefont
  {Creton}}\ and\ \bibinfo {author} {\bibfnamefont {M.}~\bibnamefont
  {Ciccotti}},\ }\bibfield  {title} {\bibinfo {title} {Fracture and adhesion of
  soft materials: A review},\ }\href
  {https://doi.org/10.1088/0034-4885/79/4/046601} {\bibfield  {journal}
  {\bibinfo  {journal} {Reports on Progress in Physics}\ }\textbf {\bibinfo
  {volume} {79}},\ \bibinfo {pages} {046601} (\bibinfo {year}
  {2016})}\BibitemShut {NoStop}%
\bibitem [{\citenamefont {Zhou}\ \emph {et~al.}(2016)\citenamefont {Zhou},
  \citenamefont {Yu}, \citenamefont {Shao}, \citenamefont {Zhang},\ and\
  \citenamefont {Wang}}]{zhou2016viscoelastic}%
  \BibitemOpen
  \bibfield  {author} {\bibinfo {author} {\bibfnamefont {X.~Q.}\ \bibnamefont
  {Zhou}}, \bibinfo {author} {\bibfnamefont {D.~Y.}\ \bibnamefont {Yu}},
  \bibinfo {author} {\bibfnamefont {X.~Y.}\ \bibnamefont {Shao}}, \bibinfo
  {author} {\bibfnamefont {S.~Q.}\ \bibnamefont {Zhang}},\ and\ \bibinfo
  {author} {\bibfnamefont {S.}~\bibnamefont {Wang}},\ }\bibfield  {title}
  {\bibinfo {title} {Research and applications of viscoelastic vibration
  damping materials: A review},\ }\href
  {https://doi.org/10.1016/j.compstruct.2015.10.014} {\bibfield  {journal}
  {\bibinfo  {journal} {Composite Structures}\ }\textbf {\bibinfo {volume}
  {136}},\ \bibinfo {pages} {460} (\bibinfo {year} {2016})}\BibitemShut
  {NoStop}%
\bibitem [{\citenamefont {Li}\ \emph {et~al.}(2022)\citenamefont {Li},
  \citenamefont {Pal}, \citenamefont {Aghakhani}, \citenamefont
  {Pena-Francesch},\ and\ \citenamefont {Sitti}}]{li2022soft}%
  \BibitemOpen
  \bibfield  {author} {\bibinfo {author} {\bibfnamefont {M.}~\bibnamefont
  {Li}}, \bibinfo {author} {\bibfnamefont {A.}~\bibnamefont {Pal}}, \bibinfo
  {author} {\bibfnamefont {A.}~\bibnamefont {Aghakhani}}, \bibinfo {author}
  {\bibfnamefont {A.}~\bibnamefont {Pena-Francesch}},\ and\ \bibinfo {author}
  {\bibfnamefont {M.}~\bibnamefont {Sitti}},\ }\bibfield  {title} {\bibinfo
  {title} {Soft actuators for real-world applications},\ }\href
  {https://doi.org/10.1038/s41578-021-00389-7} {\bibfield  {journal} {\bibinfo
  {journal} {Nature Reviews Materials}\ }\textbf {\bibinfo {volume} {7}},\
  \bibinfo {pages} {235} (\bibinfo {year} {2022})}\BibitemShut {NoStop}%
\bibitem [{\citenamefont {Paul}\ and\ \citenamefont
  {Pusey}(1981)}]{paul1981observation}%
  \BibitemOpen
  \bibfield  {author} {\bibinfo {author} {\bibfnamefont {G.~L.}\ \bibnamefont
  {Paul}}\ and\ \bibinfo {author} {\bibfnamefont {P.~N.}\ \bibnamefont
  {Pusey}},\ }\bibfield  {title} {\bibinfo {title} {Observation of a long-time
  tail in {B}rownian motion},\ }\href
  {https://doi.org/10.1088/0305-4470/14/12/025} {\bibfield  {journal} {\bibinfo
   {journal} {Journal of Physics A: Mathematical and General}\ }\textbf
  {\bibinfo {volume} {14}},\ \bibinfo {pages} {3301} (\bibinfo {year}
  {1981})}\BibitemShut {NoStop}%
\bibitem [{\citenamefont {Cichocki}\ and\ \citenamefont
  {Felderhof}(2000)}]{cichocki2000long}%
  \BibitemOpen
  \bibfield  {author} {\bibinfo {author} {\bibfnamefont {B.}~\bibnamefont
  {Cichocki}}\ and\ \bibinfo {author} {\bibfnamefont {B.~U.}\ \bibnamefont
  {Felderhof}},\ }\bibfield  {title} {\bibinfo {title} {Long-time tails in the
  solid-body motion of a sphere immersed in a suspension},\ }\href
  {https://doi.org/10.1103/PhysRevE.62.5383} {\bibfield  {journal} {\bibinfo
  {journal} {Physical Review E}\ }\textbf {\bibinfo {volume} {62}},\ \bibinfo
  {pages} {5383} (\bibinfo {year} {2000})}\BibitemShut {NoStop}%
\bibitem [{\citenamefont {Franosch}\ \emph {et~al.}(2011)\citenamefont
  {Franosch}, \citenamefont {Grimm}, \citenamefont {Belushkin}, \citenamefont
  {Mor}, \citenamefont {Foffi}, \citenamefont {Forr{'o}},\ and\ \citenamefont
  {Jeney}}]{franosch2011resonances}%
  \BibitemOpen
  \bibfield  {author} {\bibinfo {author} {\bibfnamefont {T.}~\bibnamefont
  {Franosch}}, \bibinfo {author} {\bibfnamefont {M.}~\bibnamefont {Grimm}},
  \bibinfo {author} {\bibfnamefont {M.}~\bibnamefont {Belushkin}}, \bibinfo
  {author} {\bibfnamefont {F.~M.}\ \bibnamefont {Mor}}, \bibinfo {author}
  {\bibfnamefont {G.}~\bibnamefont {Foffi}}, \bibinfo {author} {\bibfnamefont
  {L.}~\bibnamefont {Forr{'o}}},\ and\ \bibinfo {author} {\bibfnamefont
  {S.}~\bibnamefont {Jeney}},\ }\bibfield  {title} {\bibinfo {title}
  {Resonances arising from hydrodynamic memory in {B}rownian motion},\ }\href
  {https://doi.org/10.1038/nature10498} {\bibfield  {journal} {\bibinfo
  {journal} {Nature}\ }\textbf {\bibinfo {volume} {478}},\ \bibinfo {pages}
  {85} (\bibinfo {year} {2011})}\BibitemShut {NoStop}%
\bibitem [{\citenamefont {Cates}\ and\ \citenamefont
  {Candau}(1990)}]{cates1990statics}%
  \BibitemOpen
  \bibfield  {author} {\bibinfo {author} {\bibfnamefont {M.~E.}\ \bibnamefont
  {Cates}}\ and\ \bibinfo {author} {\bibfnamefont {S.~J.}\ \bibnamefont
  {Candau}},\ }\bibfield  {title} {\bibinfo {title} {Statics and dynamics of
  worm-like surfactant micelles},\ }\href
  {https://doi.org/10.1088/0953-8984/2/33/001} {\bibfield  {journal} {\bibinfo
  {journal} {Journal of Physics: Condensed Matter}\ }\textbf {\bibinfo {volume}
  {2}},\ \bibinfo {pages} {6869} (\bibinfo {year} {1990})}\BibitemShut
  {NoStop}%
\bibitem [{\citenamefont {Baiesi}\ \emph {et~al.}(2021)\citenamefont {Baiesi},
  \citenamefont {Iubini},\ and\ \citenamefont {Orlandini}}]{baiesi2021rise}%
  \BibitemOpen
  \bibfield  {author} {\bibinfo {author} {\bibfnamefont {M.}~\bibnamefont
  {Baiesi}}, \bibinfo {author} {\bibfnamefont {S.}~\bibnamefont {Iubini}},\
  and\ \bibinfo {author} {\bibfnamefont {E.}~\bibnamefont {Orlandini}},\
  }\bibfield  {title} {\bibinfo {title} {The rise and fall of branching: A
  slowing down mechanism in relaxing wormlike micellar networks},\ }\href
  {https://doi.org/10.1063/5.0072374} {\bibfield  {journal} {\bibinfo
  {journal} {The Journal of Chemical Physics}\ }\textbf {\bibinfo {volume}
  {155}},\ \bibinfo {pages} {214905} (\bibinfo {year} {2021})}\BibitemShut
  {NoStop}%
\bibitem [{\citenamefont {Sung}\ \emph {et~al.}(2003)\citenamefont {Sung},
  \citenamefont {Han},\ and\ \citenamefont {Kim}}]{sung2003rheological}%
  \BibitemOpen
  \bibfield  {author} {\bibinfo {author} {\bibfnamefont {K.}~\bibnamefont
  {Sung}}, \bibinfo {author} {\bibfnamefont {M.-S.}\ \bibnamefont {Han}},\ and\
  \bibinfo {author} {\bibfnamefont {C.}~\bibnamefont {Kim}},\ }\bibfield
  {title} {\bibinfo {title} {Rheological behavior and wall slip of dilute and
  semidilute {CP}y{C}l/{N}a{S}al surfactant solutions},\ }\href
  {https://koreascience.kr/article/JAKO200311921923956.pdf} {\bibfield
  {journal} {\bibinfo  {journal} {Korea-Australia Rheology Journal}\ }\textbf
  {\bibinfo {volume} {15}},\ \bibinfo {pages} {151} (\bibinfo {year}
  {2003})}\BibitemShut {NoStop}%
\bibitem [{\citenamefont {Spenley}\ \emph {et~al.}(1993)\citenamefont
  {Spenley}, \citenamefont {Cates},\ and\ \citenamefont
  {McLeish}}]{SpenleyCates1993}%
  \BibitemOpen
  \bibfield  {author} {\bibinfo {author} {\bibfnamefont {N.~A.}\ \bibnamefont
  {Spenley}}, \bibinfo {author} {\bibfnamefont {M.~E.}\ \bibnamefont {Cates}},\
  and\ \bibinfo {author} {\bibfnamefont {T.~C.~B.}\ \bibnamefont {McLeish}},\
  }\bibfield  {title} {\bibinfo {title} {Nonlinear rheology of wormlike
  micelles},\ }\href {https://doi.org/10.1103/PhysRevLett.71.939} {\bibfield
  {journal} {\bibinfo  {journal} {Phys. Rev. Lett.}\ }\textbf {\bibinfo
  {volume} {71}},\ \bibinfo {pages} {939} (\bibinfo {year} {1993})}\BibitemShut
  {NoStop}%
\bibitem [{\citenamefont {Gomez-Solano}\ and\ \citenamefont
  {Bechinger}(2014)}]{gomez2014probing}%
  \BibitemOpen
  \bibfield  {author} {\bibinfo {author} {\bibfnamefont {J.~R.}\ \bibnamefont
  {Gomez-Solano}}\ and\ \bibinfo {author} {\bibfnamefont {C.}~\bibnamefont
  {Bechinger}},\ }\bibfield  {title} {\bibinfo {title} {Probing linear and
  nonlinear microrheology of viscoelastic fluids},\ }\href
  {https://doi.org/10.1209/0295-5075/108/54008} {\bibfield  {journal} {\bibinfo
   {journal} {Europhysics Letters}\ }\textbf {\bibinfo {volume} {108}},\
  \bibinfo {pages} {54008} (\bibinfo {year} {2014})}\BibitemShut {NoStop}%
\bibitem [{\citenamefont {Caspers}\ \emph {et~al.}(2023)\citenamefont
  {Caspers}, \citenamefont {Ditz}, \citenamefont {Krishna~Kumar}, \citenamefont
  {Ginot}, \citenamefont {Bechinger}, \citenamefont {Fuchs},\ and\
  \citenamefont {Krüger}}]{caspers2023mobility}%
  \BibitemOpen
  \bibfield  {author} {\bibinfo {author} {\bibfnamefont {J.}~\bibnamefont
  {Caspers}}, \bibinfo {author} {\bibfnamefont {N.}~\bibnamefont {Ditz}},
  \bibinfo {author} {\bibfnamefont {K.}~\bibnamefont {Krishna~Kumar}}, \bibinfo
  {author} {\bibfnamefont {F.}~\bibnamefont {Ginot}}, \bibinfo {author}
  {\bibfnamefont {C.}~\bibnamefont {Bechinger}}, \bibinfo {author}
  {\bibfnamefont {M.}~\bibnamefont {Fuchs}},\ and\ \bibinfo {author}
  {\bibfnamefont {M.}~\bibnamefont {Krüger}},\ }\bibfield  {title} {\bibinfo
  {title} {How are mobility and friction related in viscoelastic fluids?},\
  }\href {https://doi.org/10.1063/5.0129639} {\bibfield  {journal} {\bibinfo
  {journal} {The Journal of Chemical Physics}\ }\textbf {\bibinfo {volume}
  {158}},\ \bibinfo {pages} {024901} (\bibinfo {year} {2023})}\BibitemShut
  {NoStop}%
\bibitem [{\citenamefont {Wilhelm}\ \emph {et~al.}(2003)\citenamefont
  {Wilhelm}, \citenamefont {Browaeys}, \citenamefont {Ponton},\ and\
  \citenamefont {Bacri}}]{wilhelm2003rotational}%
  \BibitemOpen
  \bibfield  {author} {\bibinfo {author} {\bibfnamefont {C.}~\bibnamefont
  {Wilhelm}}, \bibinfo {author} {\bibfnamefont {J.}~\bibnamefont {Browaeys}},
  \bibinfo {author} {\bibfnamefont {A.}~\bibnamefont {Ponton}},\ and\ \bibinfo
  {author} {\bibfnamefont {J.-C.}\ \bibnamefont {Bacri}},\ }\bibfield  {title}
  {\bibinfo {title} {Rotational magnetic particles microrheology: {T}he
  {M}axwellian case},\ }\href {https://doi.org/10.1103/PhysRevE.67.011504}
  {\bibfield  {journal} {\bibinfo  {journal} {Phys. Rev. E}\ }\textbf {\bibinfo
  {volume} {67}},\ \bibinfo {pages} {011504} (\bibinfo {year}
  {2003})}\BibitemShut {NoStop}%
\bibitem [{\citenamefont {Janssen}\ \emph {et~al.}(2009)\citenamefont
  {Janssen}, \citenamefont {Schellekens}, \citenamefont {van Ommering},
  \citenamefont {van IJzendoorn},\ and\ \citenamefont
  {Prins}}]{janssen2009controlled}%
  \BibitemOpen
  \bibfield  {author} {\bibinfo {author} {\bibfnamefont {X.~J.~A.}\
  \bibnamefont {Janssen}}, \bibinfo {author} {\bibfnamefont {A.~J.}\
  \bibnamefont {Schellekens}}, \bibinfo {author} {\bibfnamefont
  {K.}~\bibnamefont {van Ommering}}, \bibinfo {author} {\bibfnamefont {L.~J.}\
  \bibnamefont {van IJzendoorn}},\ and\ \bibinfo {author} {\bibfnamefont
  {M.~W.~J.}\ \bibnamefont {Prins}},\ }\bibfield  {title} {\bibinfo {title}
  {Controlled torque on superparamagnetic beads for functional biosensors},\
  }\href {https://doi.org/10.1016/j.bios.2008.09.024} {\bibfield  {journal}
  {\bibinfo  {journal} {Biosensors and Bioelectronics}\ }\textbf {\bibinfo
  {volume} {24}},\ \bibinfo {pages} {1937} (\bibinfo {year}
  {2009})}\BibitemShut {NoStop}%
\bibitem [{\citenamefont {Rehage}\ and\ \citenamefont
  {Hoffmann}(1998)}]{rehage1988rheological}%
  \BibitemOpen
  \bibfield  {author} {\bibinfo {author} {\bibfnamefont {H.}~\bibnamefont
  {Rehage}}\ and\ \bibinfo {author} {\bibfnamefont {H.}~\bibnamefont
  {Hoffmann}},\ }\bibfield  {title} {\bibinfo {title} {Rheological properties
  of viscoelastic surfactant systems},\ }\href
  {https://doi.org/10.1021/j100327a031} {\bibfield  {journal} {\bibinfo
  {journal} {The Journal of Physical Chemistry}\ }\textbf {\bibinfo {volume}
  {92}},\ \bibinfo {pages} {4712} (\bibinfo {year} {1998})}\BibitemShut
  {NoStop}%
\bibitem [{\citenamefont {Ginot}\ \emph {et~al.}(2022)\citenamefont {Ginot},
  \citenamefont {Caspers}, \citenamefont {Reinalter}, \citenamefont {{Krishna
  Kumar}}, \citenamefont {Krüger},\ and\ \citenamefont
  {Bechinger}}]{ginot2022recoil}%
  \BibitemOpen
  \bibfield  {author} {\bibinfo {author} {\bibfnamefont {F.}~\bibnamefont
  {Ginot}}, \bibinfo {author} {\bibfnamefont {J.}~\bibnamefont {Caspers}},
  \bibinfo {author} {\bibfnamefont {L.~F.}\ \bibnamefont {Reinalter}}, \bibinfo
  {author} {\bibfnamefont {K.}~\bibnamefont {{Krishna Kumar}}}, \bibinfo
  {author} {\bibfnamefont {M.}~\bibnamefont {Krüger}},\ and\ \bibinfo {author}
  {\bibfnamefont {C.}~\bibnamefont {Bechinger}},\ }\bibfield  {title} {\bibinfo
  {title} {Recoil experiments determine the eigenmodes of viscoelastic
  fluids},\ }\href {https://doi.org/10.1088/1367-2630/aca8c7} {\bibfield
  {journal} {\bibinfo  {journal} {New Journal of Physics}\ }\textbf {\bibinfo
  {volume} {24}},\ \bibinfo {pages} {123013} (\bibinfo {year}
  {2022})}\BibitemShut {NoStop}%
\bibitem [{\citenamefont {Shao-hua}(2004)}]{Guo2004EigenViscoelastic}%
  \BibitemOpen
  \bibfield  {author} {\bibinfo {author} {\bibfnamefont {G.}~\bibnamefont
  {Shao-hua}},\ }\bibfield  {title} {\bibinfo {title} {Eigen theory of
  viscoelastic dynamics based on the {K}elvin--{V}oigt model},\ }\href
  {https://doi.org/10.1007/BF02437571} {\bibfield  {journal} {\bibinfo
  {journal} {Applied Mathematics and Mechanics (English Edition)}\ }\textbf
  {\bibinfo {volume} {25}},\ \bibinfo {pages} {792} (\bibinfo {year}
  {2004})}\BibitemShut {NoStop}%
\bibitem [{\citenamefont {Serra-Aguila}\ \emph {et~al.}(2019)\citenamefont
  {Serra-Aguila}, \citenamefont {Puigoriol-Forcada}, \citenamefont {Reyes},\
  and\ \citenamefont {Menacho}}]{SerraAguila2019}%
  \BibitemOpen
  \bibfield  {author} {\bibinfo {author} {\bibfnamefont {A.}~\bibnamefont
  {Serra-Aguila}}, \bibinfo {author} {\bibfnamefont {J.~M.}\ \bibnamefont
  {Puigoriol-Forcada}}, \bibinfo {author} {\bibfnamefont {G.}~\bibnamefont
  {Reyes}},\ and\ \bibinfo {author} {\bibfnamefont {J.}~\bibnamefont
  {Menacho}},\ }\bibfield  {title} {\bibinfo {title} {Viscoelastic models
  revisited: characteristics and interconversion formulas for generalized
  {K}elvin--{V}oigt and {M}axwell models},\ }\href
  {https://doi.org/10.1007/s10409-019-00895-6} {\bibfield  {journal} {\bibinfo
  {journal} {Acta Mechanica Sinica}\ }\textbf {\bibinfo {volume} {35}},\
  \bibinfo {pages} {1191} (\bibinfo {year} {2019})}\BibitemShut {NoStop}%
\bibitem [{\citenamefont {Saha}\ and\ \citenamefont
  {Krüger}(2026)}]{saha2026peculiar}%
  \BibitemOpen
  \bibfield  {author} {\bibinfo {author} {\bibfnamefont {R.}~\bibnamefont
  {Saha}}\ and\ \bibinfo {author} {\bibfnamefont {M.}~\bibnamefont {Krüger}},\
  }\bibfield  {title} {\bibinfo {title} {The peculiar response of
  {K}elvin–{V}oigt chains with a free end},\ }\href
  {https://doi.org/10.1088/1751-8121/ae8b46} {\bibfield  {journal} {\bibinfo
  {journal} {Journal of Physics A: Mathematical and Theoretical}\ }\textbf
  {\bibinfo {volume} {59}},\ \bibinfo {pages} {315003} (\bibinfo {year}
  {2026})}\BibitemShut {NoStop}%
\bibitem [{Note1()}]{Note1}%
  \BibitemOpen
  \bibinfo {note} {The boundary conditions turn into a free boundary condition
  at $r=R$ and a Dirichlet boundary condition at $r=L$. $\Delta r$ is the
  radial distance between neighboring shells.}\BibitemShut {Stop}%
\bibitem [{\citenamefont {Landau}\ and\ \citenamefont
  {Lifshitz}(1987)}]{landau1987fluid}%
  \BibitemOpen
  \bibfield  {author} {\bibinfo {author} {\bibfnamefont {L.~D.}\ \bibnamefont
  {Landau}}\ and\ \bibinfo {author} {\bibfnamefont {E.~M.}\ \bibnamefont
  {Lifshitz}},\ }\emph {\bibinfo {title} {Fluid Mechanics:
  Volume 6}},\ Vol.~\bibinfo {volume} {6}\ (\bibinfo  {publisher} {Elsevier},\
  \bibinfo {year} {1987})\BibitemShut {NoStop}%
\bibitem [{Note2()}]{Note2}%
  \BibitemOpen
  \bibinfo {note} {This is because arc length is radius multiplied by covered
  angle.}\BibitemShut {Stop}%
\bibitem [{\citenamefont {Wilking}\ and\ \citenamefont
  {Mason}(2008)}]{wilking2008optically}%
  \BibitemOpen
  \bibfield  {author} {\bibinfo {author} {\bibfnamefont {J.~N.}\ \bibnamefont
  {Wilking}}\ and\ \bibinfo {author} {\bibfnamefont {T.~G.}\ \bibnamefont
  {Mason}},\ }\bibfield  {title} {\bibinfo {title} {Optically driven nonlinear
  microrheology of gelatin},\ }\href
  {https://doi.org/10.1103/PhysRevE.77.055101} {\bibfield  {journal} {\bibinfo
  {journal} {Phys. Rev. E}\ }\textbf {\bibinfo {volume} {77}},\ \bibinfo
  {pages} {055101(R)} (\bibinfo {year} {2008})}\BibitemShut {NoStop}%
\bibitem [{\citenamefont {Habdas}\ and\ \citenamefont
  {Weeks}(2025)}]{habdas2025stirring}%
  \BibitemOpen
  \bibfield  {author} {\bibinfo {author} {\bibfnamefont {P.}~\bibnamefont
  {Habdas}}\ and\ \bibinfo {author} {\bibfnamefont {E.~R.}\ \bibnamefont
  {Weeks}},\ }\bibfield  {title} {\bibinfo {title} {Stirring supercooled
  colloidal liquids at the particle scale},\ }\href
  {https://doi.org/10.1103/j1q2-h4wz} {\bibfield  {journal} {\bibinfo
  {journal} {Phys. Rev. E}\ }\textbf {\bibinfo {volume} {111}},\ \bibinfo
  {pages} {065415} (\bibinfo {year} {2025})}\BibitemShut {NoStop}%
\bibitem [{\citenamefont {Hertlein}\ \emph {et~al.}(2008)\citenamefont
  {Hertlein}, \citenamefont {Helden}, \citenamefont {Gambassi}, \citenamefont
  {Dietrich},\ and\ \citenamefont {Bechinger}}]{hertlein2008direct}%
  \BibitemOpen
  \bibfield  {author} {\bibinfo {author} {\bibfnamefont {C.}~\bibnamefont
  {Hertlein}}, \bibinfo {author} {\bibfnamefont {L.}~\bibnamefont {Helden}},
  \bibinfo {author} {\bibfnamefont {A.}~\bibnamefont {Gambassi}}, \bibinfo
  {author} {\bibfnamefont {S.}~\bibnamefont {Dietrich}},\ and\ \bibinfo
  {author} {\bibfnamefont {C.}~\bibnamefont {Bechinger}},\ }\bibfield  {title}
  {\bibinfo {title} {Direct measurement of critical {C}asimir forces},\ }\href
  {https://doi.org/10.1038/nature06443} {\bibfield  {journal} {\bibinfo
  {journal} {Nature}\ }\textbf {\bibinfo {volume} {451}},\ \bibinfo {pages}
  {172} (\bibinfo {year} {2008})}\BibitemShut {NoStop}%
\bibitem [{\citenamefont {Pi\ifmmode~\check{s}\else \v{s}\fi{}ljar}\ \emph
  {et~al.}(2022)\citenamefont {Pi\ifmmode~\check{s}\else \v{s}\fi{}ljar},
  \citenamefont {Ghosh}, \citenamefont {Turlapati}, \citenamefont {Rao},
  \citenamefont {\ifmmode~\check{S}\else \v{S}\fi{}karabot}, \citenamefont
  {Mertelj}, \citenamefont {Petelin}, \citenamefont {Nych}, \citenamefont
  {Marin\ifmmode \check{c}\else \v{c}\fi{}i\ifmmode~\check{c}\else \v{c}\fi{}},
  \citenamefont {Pusovnik}, \citenamefont {Ravnik},\ and\ \citenamefont
  {Mu\ifmmode \check{s}\else \v{s}\fi{}evi\ifmmode~\check{c}\else
  \v{c}\fi{}}}]{pivsljar2022blue}%
  \BibitemOpen
  \bibfield  {author} {\bibinfo {author} {\bibfnamefont {J.}~\bibnamefont
  {Pi\ifmmode~\check{s}\else \v{s}\fi{}ljar}}, \bibinfo {author} {\bibfnamefont
  {S.}~\bibnamefont {Ghosh}}, \bibinfo {author} {\bibfnamefont
  {S.}~\bibnamefont {Turlapati}}, \bibinfo {author} {\bibfnamefont {N.~V.~S.}\
  \bibnamefont {Rao}}, \bibinfo {author} {\bibfnamefont {M.}~\bibnamefont
  {\ifmmode~\check{S}\else \v{S}\fi{}karabot}}, \bibinfo {author}
  {\bibfnamefont {A.}~\bibnamefont {Mertelj}}, \bibinfo {author} {\bibfnamefont
  {A.}~\bibnamefont {Petelin}}, \bibinfo {author} {\bibfnamefont
  {A.}~\bibnamefont {Nych}}, \bibinfo {author} {\bibfnamefont {M.}~\bibnamefont
  {Marin\ifmmode \check{c}\else \v{c}\fi{}i\ifmmode~\check{c}\else
  \v{c}\fi{}}}, \bibinfo {author} {\bibfnamefont {A.}~\bibnamefont {Pusovnik}},
  \bibinfo {author} {\bibfnamefont {M.}~\bibnamefont {Ravnik}},\ and\ \bibinfo
  {author} {\bibfnamefont {I.}~\bibnamefont {Mu\ifmmode \check{s}\else
  \v{s}\fi{}evi\ifmmode~\check{c}\else \v{c}\fi{}}},\ }\bibfield  {title}
  {\bibinfo {title} {Blue {P}hase {III}: {T}opological {F}luid of
  {S}kyrmions},\ }\href {https://doi.org/10.1103/PhysRevX.12.011003} {\bibfield
   {journal} {\bibinfo  {journal} {Phys. Rev. X}\ }\textbf {\bibinfo {volume}
  {12}},\ \bibinfo {pages} {011003} (\bibinfo {year} {2022})}\BibitemShut
  {NoStop}%
\bibitem [{\citenamefont {Guo}\ \emph {et~al.}(2014)\citenamefont {Guo},
  \citenamefont {Ehrlicher}, \citenamefont {Jensen}, \citenamefont {Renz},
  \citenamefont {Moore}, \citenamefont {Goldman}, \citenamefont
  {Lippincott-Schwartz}, \citenamefont {Mackintosh},\ and\ \citenamefont
  {Weitz}}]{guo2014probing}%
  \BibitemOpen
  \bibfield  {author} {\bibinfo {author} {\bibfnamefont {M.}~\bibnamefont
  {Guo}}, \bibinfo {author} {\bibfnamefont {A.~J.}\ \bibnamefont {Ehrlicher}},
  \bibinfo {author} {\bibfnamefont {M.~H.}\ \bibnamefont {Jensen}}, \bibinfo
  {author} {\bibfnamefont {M.}~\bibnamefont {Renz}}, \bibinfo {author}
  {\bibfnamefont {J.~R.}\ \bibnamefont {Moore}}, \bibinfo {author}
  {\bibfnamefont {R.~D.}\ \bibnamefont {Goldman}}, \bibinfo {author}
  {\bibfnamefont {J.}~\bibnamefont {Lippincott-Schwartz}}, \bibinfo {author}
  {\bibfnamefont {F.~C.}\ \bibnamefont {Mackintosh}},\ and\ \bibinfo {author}
  {\bibfnamefont {D.~A.}\ \bibnamefont {Weitz}},\ }\bibfield  {title} {\bibinfo
  {title} {Probing the {S}tochastic, {M}otor-{D}riven {P}roperties of the
  {C}ytoplasm {U}sing {F}orce {S}pectrum {M}icroscopy},\ }\href
  {https://doi.org/10.1016/j.cell.2014.06.051} {\bibfield  {journal} {\bibinfo
  {journal} {Cell}\ }\textbf {\bibinfo {volume} {158}},\ \bibinfo {pages}
  {822–832} (\bibinfo {year} {2014})}\BibitemShut {NoStop}%
\bibitem [{\citenamefont {Kos}\ and\ \citenamefont
  {Dunkel}(2022)}]{kos2022nematic}%
  \BibitemOpen
  \bibfield  {author} {\bibinfo {author} {\bibfnamefont {{\v Z}.}~\bibnamefont
  {Kos}}\ and\ \bibinfo {author} {\bibfnamefont {J.}~\bibnamefont {Dunkel}},\
  }\bibfield  {title} {\bibinfo {title} {Nematic bits and universal logic
  gates},\ }\href {https://doi.org/10.1126/sciadv.abp8371} {\bibfield
  {journal} {\bibinfo  {journal} {Science Advances}\ }\textbf {\bibinfo
  {volume} {8}},\ \bibinfo {pages} {eabp8371} (\bibinfo {year}
  {2022})}\BibitemShut {NoStop}%
\bibitem [{\citenamefont {Woodhouse}\ and\ \citenamefont
  {Dunkel}(2017)}]{woodhouse2017active}%
  \BibitemOpen
  \bibfield  {author} {\bibinfo {author} {\bibfnamefont {F.~G.}\ \bibnamefont
  {Woodhouse}}\ and\ \bibinfo {author} {\bibfnamefont {J.}~\bibnamefont
  {Dunkel}},\ }\bibfield  {title} {\bibinfo {title} {Active matter logic for
  autonomous microfluidics},\ }\href {https://doi.org/10.1038/ncomms15169}
  {\bibfield  {journal} {\bibinfo  {journal} {Nature Communications}\ }\textbf
  {\bibinfo {volume} {8}},\ \bibinfo {pages} {15169} (\bibinfo {year}
  {2017})}\BibitemShut {NoStop}%
\bibitem [{\citenamefont {DeCamp}\ \emph {et~al.}(2015)\citenamefont {DeCamp},
  \citenamefont {Redner}, \citenamefont {Baskaran}, \citenamefont {Hagan},\
  and\ \citenamefont {Dogic}}]{decamp2015orientational}%
  \BibitemOpen
  \bibfield  {author} {\bibinfo {author} {\bibfnamefont {S.~J.}\ \bibnamefont
  {DeCamp}}, \bibinfo {author} {\bibfnamefont {G.~S.}\ \bibnamefont {Redner}},
  \bibinfo {author} {\bibfnamefont {A.}~\bibnamefont {Baskaran}}, \bibinfo
  {author} {\bibfnamefont {M.~F.}\ \bibnamefont {Hagan}},\ and\ \bibinfo
  {author} {\bibfnamefont {Z.}~\bibnamefont {Dogic}},\ }\bibfield  {title}
  {\bibinfo {title} {Orientational order of motile defects in active
  nematics},\ }\href {https://doi.org/10.1038/nmat4387} {\bibfield  {journal}
  {\bibinfo  {journal} {Nature Materials}\ }\textbf {\bibinfo {volume} {14}},\
  \bibinfo {pages} {1110} (\bibinfo {year} {2015})}\BibitemShut {NoStop}%
\bibitem [{\citenamefont {Roy}(2026)}]{roy2026geometry}%
  \BibitemOpen
  \bibfield  {author} {\bibinfo {author} {\bibfnamefont {N.}~\bibnamefont
  {Roy}},\ }\bibfield  {title} {\bibinfo {title} {Data for geometry-controlled
  relaxation spectra in viscoelastic fluids},\ }\href
  {https://doi.org/10.5281/zenodo.22744376} {10.5281/zenodo.22744376} (\bibinfo
  {year} {2026})\BibitemShut {NoStop}%
\end{thebibliography}
\end{document}